\documentclass[12pt]{article}

\usepackage{amssymb}
\usepackage[mathscr]{euscript}

\newtheorem{prop}{Proposition}
\newtheorem{rema}{Remark}
\newtheorem{defi}{Definition}
\newtheorem{lemm}{Lemma}
\newtheorem{theo}{Theorem}

\newcommand{\bbox}{\normalsize {}%
        \nolinebreak \hfill $\blacksquare$ \medbreak \par}

\newcommand{\inn}{\hspace*{2pt}\raisebox{-1pt}{\rule{6pt}{.3pt}\hspace*
{0pt}\rule{.3pt}{8pt}\hspace*{3pt}}}

\def\<{\langle} \def\>{\rangle}

\title{Hamiltonian formalism with several variables\\and quantum field theory\footnote{Keywords: Hamiltonian formalism, field theory, Legendre correspondance, pataplectic form, Cartan-Poincar\'e form, Poisson bracket, Interacting scalar field, conformal string theory} :\\ Part I}
\author{  
Fr\'ed\'eric H\'ELEIN\footnote{helein@cmla.ens-cachan.fr} \\
 CMLA, ENS de Cachan \\
 61,avenue du Pr\'esident Wilson\\
 94235 Cachan Cedex, France 
\\ \\
Joseph KOUNEIHER\footnote{kouneiher@paris7.jussieu.fr}\\
Universit\'e Diderot-Paris 7 \\
   case 7064\\
 75005 Paris, France  \\ 
\& \\
CNRS-URA 2052 (C.E.A.)\\
C.E. Saclay
91191 Gif-sur-Yvette Cedex}

\date{April 7, 2000}

\begin{document}
\maketitle
\abstract{We discuss in this paper the canonical structure of classical field theory in finite dimensions within the {\it{pataplectic}} Hamiltonian formulation, where we put forward the role of Legendre correspondance. We define the Poisson $\mathfrak{p}$-brackets and $\mathfrak{\omega}$-brackets which are the analogues of the Poisson bracket on forms. We formulate the equations of motion of forms in terms of $\mathfrak{p}$-brackets and $\mathfrak{\omega}$-brackets with the $n$-form ${\cal H}\omega $.  As illustration of our formalism we present two examples: the interacting scalar fields and conformal string theory.} 

\section{Introduction}
A crucial step in the formulation of Hamiltonian mechanics is the construction of the Poisson bracket between a pair of physical observables. This is obtained from the natural symplectic structure on $T^{\star}M$ (where $M$ is the configuration space of the physical system). In this phase space approache to classical mechanics, the dynamical evolution from an initial point
$x_{O} \in T^{\star}M$ is the solution to Hamilton's first order differential equations. 
Geometrically, dynamical paths in phase space can be identified with the flow lines of a special vector field $\xi_{H}$ on $T^{\star}M$ associated with the Hamiltonian fonction $H$.
Those dynamical equations imply the time rate of change of any physical observable
$f \in C^{\infty}(T^{\star}M, \Bbb {R})$, precisely through the Poisson bracket of $f$
with $H$ which is defined thanks to a Hamiltonian vector field $\xi_{f}$ on $T^{\star}M$
associated with $f$. The association of $\xi_{f}$ with $f$ means that every observable may viewed as the generator of "infinitesimal transformation" of  $T^{\star}M$. Thus every $f$ generates a
local one-parameter group of canonical transformations which is defined globally if $\xi_{f}$
is complete. Notice however that
this property depends on the topological structure of $T^{\star}M$ and is only true
in general if
the first real cohomology group $H^{1}(T^{\star}M, \Bbb{R})$ vanishes. If
$H^{1}(T^{\star}M, \Bbb{R})$ is non trivial we have vector fields which are only "locally Hamiltonian\footnote{Notice that for a large class of physically interesting systems, the dynamical vector fields are globally Hamiltonian. That is, time evolution of physically interesting systems can be generally specified simply by fixing a function $H$ on $T^{\star}M$}", and their occurence is one of the topological hazards that have to be surmounted in the quantization programme.\\ \\

In the canonical approach to a  standard field theory, the canonical variables are defined on space like hypersurfaces\footnote{In a finite dimensional classical system, the motivation fro choosing the cotangent bundle as a mathematical model for phase space lay in the possibility of identifying elements of $T^{\star}M$ with initial data for the dynamical evolution. Analogously, in a field theory we would expect the state space to consist of all Cauchy data for the system under  consideration and it is this requirement that should determine our choice of a mathematical model}. All the points on such a surface are at {\it{equal time}} and the dynamical equations specify how the canonical variables evolve from one equal time hypersurface to another, so we have an instantaneous Hamiltonian formalism on a infinite dimensional phase space\footnote{Notice that space and time are  treated asymetrically, and thus we have a non  covariance scheme .}.  More generally, let ${\cal{X}}$ and ${\cal{Y}}$ be two differentiable manifolds. From this viewpoint  a field is $u : \Bbb{R} \times {\cal{X}} \rightarrow {\cal{Y}}$. The set $C = \{ x \in {\cal{X}} \rightarrow y \in {\cal{Y}}\}$ form a  "generalized space" (the configuration sapce) on which we construct formally a cotangent bundle. \\  

Some remraks are in order:\\

\begin{itemize}
\item{ The general theory of infinite dimensional nonlinear Hamiltonian systems proceeds as in the finite dimensional case. However, there are technical difficulties related to question like the differentiability of the flow. These are outgrowths of the fact that the vector fields are densely defined, since we are dealing with partial rather than ordinary differential equations}
\item {If ${\cal{X}}$ and ${\cal{Y}}$ are any two finite dimensional $C^{\infty}$-manifolds the  simplest topology to put on  $C^{r}({\cal{X}}, {\cal{Y}})$, $r \prec \infty$, is the compact open topology in which all derivatives up to, and including, order $r$ are uniformly bounded on compact susbsets of ${\cal{X}}$. The derivatives are defined using local coordinates on  both ${\cal{X}}$ and ${\cal{Y}}$ but the topology is independent of how these are choosen. If ${\cal{X}}$ is compact,  $C^{r}({\cal{X}}, {\cal{Y}})$ has the pleasant property of being a Banach manifold modelled on its tangent spaces which are the Banach spaces $C^{r}({\cal{X}}, {\Bbb{R}}^{n})$. This property is lost when ${\cal{X}}$ is non-compact}
\item {The space $C^{\infty}({\cal{X}}, {\cal{Y}})$ may readily topologized by controlling the behavior (on compact sets) of arbitrary, but finite, orders of derivatives. However, even when ${\cal{X}}$ is compact, this is not a Banach manifold. In fact the tangent space $C^{\infty}({\cal{X}}, {\Bbb{R}}^{n})$ has the structure of a Fr\'echet space which, from the viewpoint of differential geometry, is far from ideal since (amongst other problems) the inverse function theorem is lost.}
\end{itemize}

In addition to the difficulties encountered in the classical (Hamiltonian) regime when treating the field theory canonically, we have others when we quantify the theory. For exemple, the Stone-Von Neuman theorem does not apply to the infinite  dimensional case, and there will be a large number of unitarily inequivalent representations of the  canonical commutations relations corresponding to inequivalent choices of the measure\footnote{In a large number of physically interesting cases in which the classical configuration is not a linear vector space, the question of the mesure become very hard indeed, and leads at once to the problem of what could be the analogue of a {\it{ distribution}} for such a non linear space?. For example, in the non linear $\sigma$-model it is not at all obvious what might be meant by a distribution valued analogue. } on the Hilbert space.\\ \\
Motivated by similar reflexions people try to formulate a finite dimensional (canonical) field theory and which treat the space and time in equal footing (symetrically) see for instance \cite{De Donder}, \cite{Caratha1}, \cite{Weyl1}, \cite{Rund}. Further details can be found in \cite{Kastrup1,Gotay1,Gotay2,Gotay3,Rund2,Helein}, and \cite{Dedecker1,Dedecker2,Gawcedzki1,Kijowski1,Goldschmidt1}. More recently,  a definition of the Poisson brackets on forms and the equations of motion of forms from De Donder-Weyl point of view was given for review see \cite{kanatchikov1, Kanatchikov2} and ~\cite{Hrabak1, Hrabak2, Enriquez1} a close point of view can be found in \cite{Enriquez2, Paufler, Giachetta} for others discussions see \cite{Sardanashvily1, Sardanashvily2, Sardanashvily3}. However,  we notice that in those works and contrary to the  $n-1$-form case where we have a natural link between Poisson brackets and dynamics, in the case of forms of arbitrary degrees the link is not clear. Therefore, our introduction of the $\omega$-bracket is a first attempt to resolve this difficulty, and  a generalized  $\mathfrak{p}$-bracket  between forms of arbitrary degrees wil be given in our forthcoming paper. \\ \\

 In this paper we exhibit a general construction of a {\it{universal}} Hamiltonian formalism and which generalized the schemes (of a manifest covariant  finite dimensional field theory) mentioned above, which explains the appelation {\it{universal}}. The main focus in this construction is on the role of Legendre correspondance, and the hypothesis concerning the generalized Legendre condition.\\
A motivation to study the {\it{universal}} Hamiltonian formalism is to apply it in the context of a integrable systems and to analyse the canonical structure of the physical theories, for instance general relativity and string theory with the aim to quantify those theories. So we have  to gain insight into the inherent structure of this approache, in particular  the appropriate generalization of the Poisson bracket.\\ \\
Our paper is organized as follow. In section (2) we establish the Hamiltonian formalism: the Euler-Lagrange equations, Legendre's correspondance (and the generalized Legendre condition), Hamilton's equations, Cartan-Poincar\'e and pataplectic forms. In section (3) we review the usual approache to quantum field theory. In section (4) we define the Poisson $\mathfrak{p}$-bracket wich give us the dynamics of a subset of $n-1$-forms: the $n-1$-generalized positions and momenta. To define the dynamics of the generalized positions and momenta which are not part of this subset, we introduce the $\omega$-bracket which induces the dynamics of forms of arbitrary degrees.  Finally in section (5) we present two examples: the interacting scalar fields and conformal string theory.

\section{Construction of the Hamiltonian formalism}
In this section we show how to build a universal Hamiltonian formalism for a $\sigma$-model variational problem involving a
Lagrangian functional depending on first derivatives. We derive it through a universal Legendre correspondance.

\subsection{Notations}
Let ${\cal X}$ and ${\cal Y}$ be two differentiable manifolds. ${\cal X}$ plays the role of the space-time
manifold and ${\cal Y}$ the target manifold. We fix some volume form $\omega$ on ${\cal X}$, this
volume form may be chosen according to the variational problem that we want to study (for instance if
we look at the Klein-Gordon functional on some pseudo-Riemannian manifold, we choose $\omega$ to be the Riemannian volume),
but in more general situation, with less symmetries we just choose some arbitrary volume form.
We set $n=\hbox{dim}{\cal X}$ and $k=\hbox{dim}{\cal Y}$. We
denote $\{ x^1,...,x^n\}$ local coordinates on ${\cal X}$ and
$\{ y^1,...,y^k\}$ local coordinates on ${\cal Y}$.
For simplicity we shall assume that the coordinates $x^{\alpha}$
are always chosen such that $dx^1\wedge ...\wedge dx^n =\omega$,
through it is not essential. Then on the product
${\cal X}\times {\cal Y}$ we denote $\{ q^1,...,q^{n+k}\}$ local coordinates in such a
way that

$$\begin{array}{cl}
q^{\mu} = x^{\mu} = x^{\alpha} & \hbox{ if } 1\leq \mu=\alpha \leq n\\
q^{\mu} = y^{\mu-n} = y^i & \hbox{ if } 1\leq \mu-n =i \leq k.
\end{array}$$
Generally we shall denote the
indices running from 1 to $n$ by $\alpha$, $\beta$,... , the indices
between 1 and $k$ by $i$, $j$, ... and the indices between 1 and $n+k$ by
$\mu$, $\nu$,...
To any map $u:{\cal X}\longrightarrow {\cal Y}$ we may associate the map

$$\begin{array}{cccc}
U: & {\cal X} & \longrightarrow & {\cal X}\times {\cal Y}\\
 & x & \longmapsto & (x,u(x))
\end{array}$$
whose image is the graph of $u$, $\{(x,u(x))/x\in {\cal X}\}$.
We also associate to $u$ the bundle $u^{\star}T{\cal Y}\otimes T^{\star}{\cal X}$ over ${\cal X}$.
This bundle is naturally equipped with the coordinates $(x^{\alpha})_{1\leq \alpha \leq n}$
(for ${\cal X}$) and $(v^i_{\alpha})_{1\leq i\leq k;1\leq \alpha \leq n}$, such that a
point $(x,v)\in u^{\star}T{\cal Y}\otimes T^{\star}{\cal X}$ is represented by

$$v = \sum_{\alpha=1}^n\sum_{i=1}^kv^i_{\alpha}{\partial \over \partial y^i}\otimes dx^{\alpha}.$$
We can think $u^{\star}T{\cal Y}\otimes T^{\star}{\cal X}$ as a subset
of $T{\cal Y}\otimes T^{\star}{\cal X}:=\{(x,y,v)/(x,y)\in {\cal X}\times {\cal Y},
v\in T_y{\cal Y}\otimes T^{\star}_x{\cal X}\}$ by the inclusion map
$(x,v)\longmapsto (x,u(x),v)$.

The differential of $u$, $du$ is a section of the bundle $u^{\star}T{\cal Y}\otimes T^{\star}{\cal X}$ over ${\cal X}$. 
Hence the coordinates for $du$ are simply $v^i_{\alpha} = {\partial u^i\over \partial x^{\alpha}}$.
Notice that $u^{\star}T{\cal Y}\otimes T^{\star}{\cal X}$ is a kind of analog of of the tangent bundle
$T{\cal Y}$ to a configuration space ${\cal Y}$ in classical particle mechanics.\\

It turns out to be more convenient to consider $\Lambda^nT({\cal X}\times {\cal Y})$ the analog of $T(\Bbb{R}\times {\cal Y})$, the tangent
bundle to a space-time, or rather $S\Lambda^nT({\cal X}\times {\cal Y})$,  the  submanifold of $\Lambda^nT({\cal X}\times {\cal Y})$, as  the analog of the subset
$ST(\Bbb{R}\times {\cal Y}):=\{(t,x;\xi^0,\vec{\xi})\in T(\Bbb{R}\times {\cal Y})/dt(\xi^0,\vec{\xi})=1\}$,
which is diffeomorphic to $\Bbb{R}\times T{\cal Y}$ by the map $(t,x,\xi)\longmapsto (t,x,\vec{\xi})$, and where:

$$S\Lambda^nT({\cal X}\times {\cal Y}):= \{ (q,z)\in  \Lambda^nT({\cal X}\times {\cal Y})/
z = z_1\wedge ...\wedge z_n, z_1,...,z_n\in T_q({\cal X}\times {\cal Y}), \omega(z_1,...,z_n)=1\}.$$

For any $(x,y)\in {\cal X}\times {\cal Y}$, the fiber $S\Lambda^nT_{(x,y)}({\cal X}\times
{\cal Y})$
can be identified with $T_y{\cal Y}\otimes T^{\star}_x{\cal X}$ by the diffeomorphism

\begin {equation}\label{identify}
\begin{array}{ccc}
T_y{\cal Y}\otimes T^{\star}_x{\cal X} & \longrightarrow &
S\Lambda^nT_{(x,y)}({\cal X}\times {\cal Y})\\
v= \sum_{\alpha=1}^n\sum_{i=1}^kv^i_{\alpha}{\partial \over \partial y^i}\otimes dx^{\alpha}
& \longmapsto &
z = z_1\wedge ...\wedge z_n,
\end{array}
\end{equation}
where for all $1\leq \beta\leq n$, $z_{\beta} = {\partial \over \partial x^{\alpha}} +
\sum_{i=1}^kv^i_{\alpha}{\partial \over \partial y^i}$.
We denote by $(z^{\mu}_{\alpha})_{1\leq \mu\leq n+k;1\leq \alpha \leq n}$ the coordinates of $z_{\alpha}$, so that
$z_{\beta} =\sum_{\mu=1}^{n+k}z^{\mu}_{\alpha}{\partial \over \partial q^{\mu}}$ (or
$z^{\beta}_{\alpha} = \delta^{\beta}_{\alpha}$ for $1\leq \beta\leq n$ and
$z^{\mu+i}_{\alpha} = v^i_{\alpha}$ for $1\leq i\leq k$). This induces an identification
$T{\cal Y}\otimes T^{\star}{\cal X}\simeq S\Lambda^nT({\cal X}\times {\cal Y})$.

Thus coordinates $(x^{\alpha},y^i,v^i_{\alpha})$ (or equivalentely $(x^{\alpha},y^i,z^{\mu}_{\alpha})$) can be thought as coordinate on
$T{\cal Y}\otimes T^{\star}{\cal X}$ or $S\Lambda^nT({\cal X}\times {\cal Y})$.

Given a Lagrangian function $L:T{\cal Y}\otimes T^{\star}{\cal X}\longmapsto \Bbb{R}$, we define the
functional

$${\cal L}[u]:= \int_{\cal X}L(x,u(x),du(x))dx.$$

\subsection{The Euler-Lagrange equations}
The critical points of the action are the maps $u:{\cal X}\longrightarrow {\cal Y}$ which are solutions
of the system of Euler-Lagrange equations 

\begin{equation}\label{EL}
{\partial \over \partial x^{\alpha}}
\left( {\partial L\over \partial v^i_{\alpha}}(x,u(x),du(x))\right)
= {\partial L\over \partial y^i}(x,u(x),du(x)).
\end{equation}

This equation implies also other equations involving the {\em stress-energy} tensor
associated to $u:{\cal X}\longrightarrow {\cal Y}$:

$$S^{\alpha}_{\beta}(x):= \delta^{\alpha}_{\beta}L(x,u(x),du(x)) -
{\partial L\over \partial v^i_{\alpha}}(x,u(x),du(x))
{\partial u^i\over \partial x^{\beta}}(x).$$
Indeed for any $u$,

$$\begin{array}{ccl}
\displaystyle {\partial S^{\alpha}_{\beta}\over \partial x^{\alpha}}(x) & = &
\displaystyle \delta^{\alpha}_{\beta}\left(
{\partial L\over \partial x^{\alpha}}(x,u,du) +
{\partial L\over \partial y^i}(x,u,du){\partial u^i\over \partial x^{\alpha}}(x) +
{\partial L\over \partial v^i_{\gamma}}(x,u,du)
{\partial ^2 u^i\over \partial x^{\alpha}\partial x^{\gamma}}(x)\right) \\
 & & \displaystyle - {\partial \over \partial x^{\alpha}}
\left( {\partial L\over \partial v^i_{\alpha}}(x,u,du)\right)
{\partial u^i\over \partial x^{\beta}}(x)
- {\partial L\over \partial v^i_{\alpha}}(x,u,du)
{\partial ^2 u^i\over \partial x^{\alpha}\partial x^{\beta}}(x)\\
 & = & \displaystyle {\partial L\over \partial x^{\beta}}(x,u,du)
- \left[ {\partial \over \partial x^{\alpha}}
\left( {\partial L\over \partial v^i_{\alpha}}(x,u(x),du(x))\right)
- {\partial L\over \partial y^i}(x,u(x),du(x))\right]
{\partial u^i\over \partial x^{\beta}}(x).
\end{array}$$
Thus we conclude that if $u$ is a solution of (\ref{EL}), then

\begin{equation}\label{stress}
{\partial S^{\alpha}_{\beta}\over \partial x^{\alpha}}(x) =
{\partial L\over \partial x^{\beta}}(x,u,du).
\end{equation}
It follows that if $L$ does not depend on $x$, then $S^{\alpha}_{\beta}$ is divergence-free
for all solutions of (\ref{EL}), a property which can be predicted by Noether's
theorem.

\subsection{The Legendre correspondance}
Let ${\cal M}:= \Lambda^nT^{\star}({\cal X}\times {\cal Y})$. Every point $(q,p)\in {\cal M}$ has coordinates
$q^{\mu}$ and $p_{\mu_1...\mu_n}$ such that $p_{\mu_1...\mu_n}$ is completely antisymmetric
in $(\mu_1,...,\mu_n)$ and

$$p = \sum_{\mu_1<...<\mu_n}p_{\mu_1...\mu_n}dq^{\mu_1}\wedge ...\wedge dq^{\mu_n}.$$
We shall define a Legendre correspondance 

$$\begin{array}{ccc}
S\Lambda^nT({\cal X}\times {\cal Y})\times \Bbb{R} & \longleftrightarrow & {\cal M}= \Lambda^nT^{\star}({\cal X}\times {\cal Y})\\
(q,v,w) & \longleftrightarrow & (q,p),
\end{array}$$
where $w\in \Bbb{R}$ is some extra parameter (its signification is not clear for the moment, $w$
is related to the possibility of fixing arbitrarely the value of some Hamiltonian). Notice that we do not name it
a transform, like in the classical theory but a correspondance, since generally there will be many possible
values of $(q,p)$ corresponding to a single value of $(q,v,w)$. But we expect that in generic situations,
there corresponds a unique $(q,v,w)$ to some $(q,p)$. This correspondance is generated by the
function

$$\begin{array}{cccc}
W: & S\Lambda^nT({\cal X}\times {\cal Y})\times {\cal M} & \longrightarrow & \Bbb{R}\\
 & (q,v,p) & \longmapsto & \langle p,v\rangle -L(q,v),
\end{array}$$
where

$$\langle p,v\rangle \simeq \langle p,z\rangle :=
\langle p,z_1\wedge ...\wedge z_n\rangle =
\sum_{\mu_1,\dots ,\mu_n}p_{\mu_1...\mu_n}z_1^{\mu_1}\dots z_n^{\mu_n}.$$

\begin{defi}
We write that $(q,v,w) \longleftrightarrow (q,p)$ if and only if
\begin{equation}\label{lw}
L(q,v) + w = \langle p,v\rangle\quad \hbox{or}\quad W(q,v,p) = w
\end{equation}
and
\begin{equation}\label{dlp}
{\partial L\over \partial v^i_{\alpha}}(q,v) =
{\partial \langle p,v\rangle \over \partial  v^i_{\alpha}} =
\left\langle p,z_1\wedge \dots \wedge z_{\alpha-1}\wedge {\partial \over \partial y^i}\wedge z_{\alpha+1}\wedge
\dots \wedge z_n\right\rangle \quad \hbox{or}\quad {\partial W\over \partial v^i_{\alpha}}(q,v,p) = 0.
\end{equation}
\end{defi}
Notice that for any $(q,v,w)\in S\Lambda^nT({\cal X}\times {\cal Y})\times \Bbb{R}$ there exist
$(q,p)\in {\cal M}$ such that $(q,v,w) \longleftrightarrow (q,p)$. This will be proven in Subsection 2.6
below. But $(q,p)$ is not unique in general.
In the following we shall need to suppose that the inverse correspondance is well-defined.\\

\noindent {\bf Hypothesis: Generalized Legendre condition} {\em There exists an open subset ${\cal O}\subset {\cal M}$ which
is non empty such that for any $(q,p)\in {\cal O}$ there exists a unique $v\in T_x{\cal Y}\otimes T_y^{\star}{\cal X}$ (or
equivalentely a unique $z\in S\Lambda^nT_q({\cal X}\times {\cal Y})$) which is a critical point of
$v\longmapsto W(q,v,p)$. We denote $v={\cal V}(q,p)$ this unique solution (or $z={\cal Z}(q,p)$). We
assume further that ${\cal V}$ is a smooth function on ${\cal O}$ (or the same for ${\cal Z}$).}\\

We now suppose that this hypothesis is true. Then we can define on ${\cal O}$ the following
Hamiltonian function

$$\begin{array}{cccc}
{\cal H}: & {\cal O} & \longrightarrow & \Bbb{R}\\
 & (q,p) & \longmapsto & \langle p,{\cal V}(q,p)\rangle - L(q,{\cal V}(q,p))
=W(q,{\cal V}(q,p),p).
\end{array}$$
We then remark that (\ref{lw}) is equivalent to $w = {\cal H}(q,p)$.

We now compute the differential of ${\cal H}$. The main point is to exploit the
condition

\begin{equation}\label{defw}
{\partial W\over \partial v^i_{\alpha}}\left( q,{\cal V}(q,p),p\right) =0
\end{equation}
(which defines ${\cal V}$).

$$\begin{array}{ccl}
d{\cal H} & = & \displaystyle \sum_{\mu}{\partial W\over \partial q^{\mu}}
\left( q,{\cal V}(q,p),p\right) dq^{\mu} +
\sum_{\mu,\nu}\sum_{\alpha}{\partial W\over \partial v^{\nu}_{\alpha}}
\left( q,{\cal V}(q,p),p\right) 
{\partial {\cal V}^{\nu}_{\alpha}\over \partial q^{\mu}}dq^{\mu}\\
 & & \displaystyle +
\sum_{\nu,\alpha}\sum_{\mu_1<...<\mu_n}
{\partial W\over \partial v^{\nu}_{\alpha}}\left( q,{\cal V}(q,p),p\right) 
{\partial {\cal V}^{\nu}_{\alpha}\over \partial p_{\mu_1...\mu_n}}dp_{\mu_1...\mu_n}\\
 & & \displaystyle + \sum_{\mu_1<...<\mu_n}
{\partial W\over \partial p_{\mu_1...\mu_n}}\left( q,{\cal V}(q,p),p\right)
dp_{\mu_1...\mu_n}\\
 & = & \displaystyle \sum_{\mu}{\partial W\over \partial q^{\mu}}
\left( q,{\cal V}(q,p),p\right) dq^{\mu} +
\sum_{\mu_1<...<\mu_n}{\partial W\over \partial p_{\mu_1...\mu_n}}
\left( q,{\cal V}(q,p),p\right) dp_{\mu_1...\mu_n}.
\end{array}$$

Now since

$${\partial W\over \partial q^{\mu}}(q,v,p) = - {\partial L\over \partial q^{\mu}}(q,v,p)$$
and

$${\partial W\over \partial p_{\mu_1...\mu_n}}(q,v,p) =
\left| \begin{array}{ccc}
z^{\mu_1}_1 & \dots & z^{\mu_1}_n\\
\vdots & & \vdots \\
z^{\mu_n}_1 & \dots & z^{\mu_n}_n\\
\end{array}\right| ,$$
we get

\begin{equation}\label{dH}
d{\cal H} = - \sum_{\mu}{\partial L\over \partial q^{\mu}}
\left( q,{\cal V}(q,p),p\right) dq^{\mu} +
\sum_{\mu_1<...<\mu_n} {\cal Z}^{\mu_1...\mu_n}_{1...n}(q,p)dp_{\mu_1...\mu_n},
\end{equation}
where

$${\cal Z}^{\mu_1...\mu_n}_{1...n}(q,p) :=
\left| \begin{array}{ccc}
{\cal Z}^{\mu_1}_1(q,p) & \dots & {\cal Z}^{\mu_1}_n(q,p)\\
\vdots & & \vdots \\
{\cal Z}^{\mu_n}_1(q,p) & \dots & {\cal Z}^{\mu_n}_n(q,p)\\
\end{array}\right| $$
are the components of the $n$-vector

$${\cal Z}_1(q,p)\wedge ...\wedge {\cal Z}_n(q,p) =
\sum_{\mu_1<...<\mu_n} {\cal Z}^{\mu_1...\mu_n}_{1...n}(q,p)
{\partial \over \partial q^{\mu_1}}\wedge \dots \wedge 
{\partial \over \partial q^{\mu_n}}.$$

\noindent To conclude let us see how the stress-energy tensor appears in
this Hamiltonian setting. We define the Hamiltonian tensor on ${\cal O}$ to be
$H(q,p) = \sum_{\alpha,\beta}H^{\alpha}_{\beta}(q,p)
{\partial \over \partial x^{\alpha}}\otimes dx^{\beta}$, with

$$H^{\alpha}_{\beta}(q,p) := 
{\partial L\over \partial v^i_{\alpha}}(q,{\cal V}(q,p)){\cal V}^i_{\beta}(q,p)
- \delta^{\alpha}_{\beta}L(q,{\cal V}(q,p)).$$
It is clear that if $(x,u(x),du(x),w)\longleftrightarrow (q,p)$ then

$$H^{\alpha}_{\beta}(q,p) = - S^{\alpha}_{\beta}(x).$$

Let us now compute $H^{\alpha}_{\beta}(q,p)$. We first use (\ref{dlp})\\

\noindent $\displaystyle \sum_i{\partial L\over \partial v^i_{\alpha}}(q,{\cal V}(q,p))
{\cal V}^i_{\beta}(q,p)$
$$\begin{array}{cl}
= & \displaystyle \sum_i
{\partial \langle p,v\rangle \over \partial  v^i_{\alpha}}_{|v={\cal V}(q,p)}
{\cal V}^i_{\beta}(q,p)\\
= & \displaystyle \sum_i
\left\langle p,{\cal Z}_1(q,p)\wedge ...\wedge {\cal Z}_{\alpha-1}(q,p)\wedge
{\partial \over \partial y^i}\wedge {\cal Z}_{\alpha+1}(q,p)\wedge ...
\wedge {\cal Z}_n(q,p)\right\rangle {\cal V}^i_{\beta}(q,p)\\
= & \displaystyle \sum_{\mu}
\left\langle p,{\cal Z}_1(q,p)\wedge ...\wedge {\cal Z}_{\alpha-1}(q,p)\wedge
{\partial \over \partial q^{\mu}}
\wedge {\cal Z}_{\alpha+1}(q,p)\wedge ...
\wedge {\cal Z}_n(q,p)\right\rangle {\cal Z}^{\mu}_{\beta}(q,p)\\
& \displaystyle
- \left\langle p,{\cal Z}_1(q,p)\wedge ...\wedge {\cal Z}_{\alpha-1}(q,p)\wedge
{\partial \over \partial x^{\beta}}\wedge {\cal Z}_{\alpha+1}(q,p)\wedge ...
\wedge {\cal Z}_n(q,p)\right\rangle \\
= & \displaystyle
\left\langle p,{\cal Z}_1(q,p)\wedge ...\wedge {\cal Z}_{\alpha-1}(q,p)\wedge
{\cal Z}_{\beta}(q,p)\wedge {\cal Z}_{\alpha+1}(q,p)\wedge ...
\wedge {\cal Z}_n(q,p)\right\rangle \\
& \displaystyle
- \left\langle p,{\cal Z}_1(q,p)\wedge ...\wedge {\cal Z}_{\alpha-1}(q,p)\wedge
{\partial \over \partial x^{\beta}}\wedge {\cal Z}_{\alpha+1}(q,p)\wedge ...
\wedge {\cal Z}_n(q,p)\right\rangle \\
= & \displaystyle
\delta^{\alpha}_{\beta}\left\langle p,{\cal Z}_1(q,p)\wedge ...
\wedge {\cal Z}_n(q,p)\right\rangle \\
& \displaystyle
- \left\langle p,{\cal Z}_1(q,p)\wedge ...\wedge {\cal Z}_{\alpha-1}(q,p)\wedge
{\partial \over \partial x^{\beta}}\wedge {\cal Z}_{\alpha+1}(q,p)\wedge ...
\wedge {\cal Z}_n(q,p)\right\rangle .
\end{array}$$
Hence since

$$\left\langle p,{\cal Z}_1(q,p)\wedge ...
\wedge {\cal Z}_n(q,p)\right\rangle = {\cal H}(q,p) + L(q,{\cal V}(q,p)),$$

\begin{equation}\label{tensh}
\begin{array}{ccl}
H^{\alpha}_{\beta}(q,p) & = & \displaystyle \delta^{\alpha}_{\beta}{\cal H}(q,p)
- \left\langle p,{\cal Z}_1(q,p)\wedge ...{\cal Z}_{\alpha-1}(q,p)\wedge
{\partial \over \partial x^{\beta}}\wedge {\cal Z}_{\alpha+1}(q,p)...
\wedge {\cal Z}_n(q,p)\right\rangle \\
 & = & \displaystyle \delta^{\alpha}_{\beta}{\cal H}(q,p)
- {\partial \langle p,z\rangle \over \partial  z^{\beta}_{\alpha}}_{|z={\cal Z}(q,p)}.
\end{array}
\end{equation}

\subsection{Hamilton equations}
Let $x\longmapsto (q(x),p(x))$ be some map from ${\cal X}$ to ${\cal O}$. To insure that this map is related to a critical point $u:{\cal X}\longrightarrow {\cal Y}$, we find that the necessary and sufficient conditions split in two parts: \\

\noindent {\bf 1) What are the conditions on $x\longmapsto (q(x),p(x))$ for the
existence of a map $x\longmapsto u(x)$ such that
$(x,u(x),du(x))\longleftrightarrow (q(x),p(x))$ ?}\\

The first obvious condition is $q(x) = (x,u(x)) = U(x)$. The second condition
is that in $T{\cal Y}\otimes T^{\star}{\cal X}$,
$(x,y,v^i_{\alpha}) = (x,y,{\partial u^i\over \partial x^{\alpha}})$ coincides with
$(q(x),{\cal V}^i_{\alpha}(q(x),p(x)))$. If we translate that using (\ref{identify}),
we obtain that in $S\Lambda^nT({\cal X}\times {\cal Y})$,

$${\partial q\over \partial x^1}\wedge \dots \wedge {\partial q\over \partial x^n} =
{\partial U\over \partial x^1}\wedge \dots \wedge {\partial U\over \partial x^n} =
{\cal Z}_1(q(x),p(x))\wedge \dots \wedge {\cal Z}_n(q(x),p(x)).$$
But we found in (\ref{dH}) that the components in the basis
$\left( {\partial \over \partial q^{\mu_1}}
\wedge \dots \wedge {\partial \over \partial q^{\mu_n}}
\right)$ of the right hand side are
${\cal Z}^{\mu_1\dots \mu_n}_{1\dots n}(q(x),p(x)) =
{\partial {\cal H}\over \partial p_{\mu_1\dots \mu_n}}(q(x),p(x))$. Hence
denoting

$${\partial (q^{\mu_1},\dots ,q^{\mu_n})\over \partial (x^1,\dots ,x^n)} :=
\left| \begin{array}{ccc}
{\partial q^{\mu_1}\over \partial x^1} & \dots & {\partial q^{\mu_1}\over \partial x^n}\\
\vdots & & \vdots \\
{\partial q^{\mu_n}\over \partial x^1} & \dots & {\partial q^{\mu_n}\over \partial x^n}
\end{array}\right| ,$$
so that

$${\partial q\over \partial x^1}\wedge \dots \wedge {\partial q\over \partial x^n} =
\sum_{\mu_1<\dots <\mu_n}
{\partial (q^{\mu_1},\dots ,q^{\mu_n})\over \partial (x^1,\dots ,x^n)}
{\partial \over \partial q^{\mu_1}}\wedge \dots \wedge {\partial \over \partial q^{\mu_n}},$$
we obtain the condition

\begin{equation}\label{hamilton1}
{\partial (q^{\mu_1},\dots ,q^{\mu_n})\over \partial (x^1,\dots ,x^n)}(x) =
{\partial {\cal H}\over \partial p_{\mu_1\dots \mu_n}}(q(x),p(x)).
\end{equation}

\noindent {\bf 2) Now what are the conditions on $x\longmapsto (q(x),p(x))$ for
$u$ to be a solution of the Euler-Lagrange equations ?}\\

It amounts to eliminate $u$ in (\ref{EL}) in function of $(q,p)$. For that
purpose we use (\ref{dlp}) to derive\\

\noindent $\displaystyle \sum_{\alpha}{\partial \over \partial x^{\alpha}}
\left( {\partial L\over \partial v^i_{\alpha}}(x,u(x),du(x))\right) $
$$\begin{array}{cl}
= &
\displaystyle \sum_{\alpha}{\partial \over \partial x^{\alpha}}
\left\langle p,{\partial U\over \partial x^1}\wedge \dots \wedge
{\partial U\over \partial x^{\alpha -1}}\wedge
{\partial \over \partial y^i}\wedge
{\partial U\over \partial x^{\alpha +1}}\wedge \dots \wedge
{\partial U\over \partial x^n}\right\rangle \\
= & \displaystyle \sum_{\alpha}
\left\langle {\partial p\over \partial x^{\alpha}},
{\partial U\over \partial x^1}\wedge \dots \wedge
{\partial U\over \partial x^{\alpha -1}}\wedge
{\partial \over \partial y^i}\wedge
{\partial U\over \partial x^{\alpha +1}}\wedge \dots \wedge
{\partial U\over \partial x^n}\right\rangle \\
& + \displaystyle \sum_{\alpha\neq \beta}
\left\langle p,{\partial U\over \partial x^1}\wedge \dots \wedge
{\partial ^2U\over \partial x^{\alpha}\partial x^{\beta}}
\wedge \dots \wedge
{\partial U\over \partial x^{\alpha -1}}\wedge
{\partial \over \partial y^i}\wedge
{\partial U\over \partial x^{\alpha +1}}\wedge \dots \wedge
{\partial U\over \partial x^n}\right\rangle \\
= & \displaystyle \sum_{\alpha}
\left\langle {\partial p\over \partial x^{\alpha}},
{\partial U\over \partial x^1}\wedge \dots \wedge
{\partial U\over \partial x^{\alpha -1}}\wedge
{\partial \over \partial y^i}\wedge
{\partial U\over \partial x^{\alpha +1}}\wedge \dots \wedge
{\partial U\over \partial x^n}\right\rangle .
\end{array}$$

On the other hand we know from (\ref{dH}) that
${\partial {\cal H}\over \partial q^i}(q,p) =
- {\partial L\over \partial q^i}(q,{\cal V}(q,p))$. Thus we obtain

\begin{equation}\label{hamilton2}
\sum_{\alpha}
\left\langle {\partial p\over \partial x^{\alpha}},
{\partial q\over \partial x^1}\wedge \dots \wedge
{\partial q\over \partial x^{\alpha -1}}\wedge
{\partial \over \partial y^i}\wedge
{\partial q\over \partial x^{\alpha +1}}\wedge \dots \wedge
{\partial q\over \partial x^n}\right\rangle =
- {\partial {\cal H}\over \partial q^i}(q(x),p(x)).
\end{equation}
The latter equation may be transformed using the relation

\noindent $\displaystyle \sum_{\alpha}
\left\langle {\partial p\over \partial x^{\alpha}},
{\partial q\over \partial x^1}\wedge \dots \wedge
{\partial q\over \partial x^{\alpha -1}}\wedge
{\partial \over \partial y^i}\wedge
{\partial q\over \partial x^{\alpha +1}}\wedge \dots \wedge
{\partial q\over \partial x^n}\right\rangle$
$$\begin{array}{cl}
= & \displaystyle
\sum_{\alpha}
\sum_{\footnotesize \begin{array}{c}\mu_1<\dots <\mu_n\\\mu_{\alpha} = n+i\end{array}}
\left| \begin{array}{ccc}
{\partial q^{\mu_1}\over \partial x^1} & \dots & {\partial q^{\mu_1}\over \partial x^n}\\
\vdots & & \vdots \\
{\partial q^{\mu_{\alpha -1}}\over \partial x^1} & \dots & {\partial q^{\mu_{\alpha -1}}\over \partial x^n}\\
{\partial p_{\mu_1\dots \mu_n}\over \partial x^1} &
\dots & {\partial p_{\mu_1\dots \mu_n}\over \partial x^n}\\
{\partial q^{\mu_{\alpha +1}}\over \partial x^1} & \dots & {\partial q^{\mu_{\alpha +1}}\over \partial x^n}\\
\vdots & & \vdots \\
{\partial q^{\mu_n}\over \partial x^1} & \dots & {\partial q^{\mu_n}\over \partial x^n}
\end{array}\right| \\
 & \\
=  & \displaystyle \sum_{\alpha}
\sum_{\footnotesize \begin{array}{c}\mu_1<\dots <\mu_n\\\mu_{\alpha} = n+i\end{array}}
{\partial (q^{\mu_1},\dots ,q^{\mu_{\alpha -1}},
p_{\mu_1\dots \mu_n}, q^{\mu_{\alpha +1}},\dots ,
q^{\mu_n})\over \partial (x^1,\dots ,x^n)}.
\end{array}$$

We summarize: the necessary and sufficient conditions we were looking for are

\begin{equation}\label{hamilton3}
\begin{array}{c}\displaystyle
{\partial (q^{\mu_1},\dots ,q^{\mu_n})\over \partial (x^1,\dots ,x^n)} =
{\partial {\cal H}\over \partial p_{\mu_1\dots \mu_n}}(q,p)\\
\\
\displaystyle
\sum_{\alpha}
\sum_{\footnotesize \begin{array}{c}\mu_1<\dots <\mu_n\\\mu_{\alpha} = n+i\end{array}}
{\partial (q^{\mu_1},\dots ,q^{\mu_{\alpha -1}},
p_{\mu_1\dots \mu_n}, q^{\mu_{\alpha +1}},\dots ,
q^{\mu_n})\over \partial (x^1,\dots ,x^n)} 
= - {\partial {\cal H}\over \partial y^i}(q,p).
\end{array}
\end{equation}

\noindent{\bf Some further relations}\\

Besides these equations, we have to remark also that equation (\ref{stress}) on the
stress-energy tensor has a counterpart in this formalism. For that purpose we
use equation (\ref{tensh}). Assuming that $(x,u(x),du(x))\longleftrightarrow
(q(x),p(x))$, we have

$$\begin{array}{ccl}
\displaystyle
- {\partial S^{\alpha}_{\beta}\over \partial x^{\alpha}}(x) & = &
\displaystyle
{\partial H^{\alpha}_{\beta}(q(x),p(x))\over \partial x^{\alpha}} \\
 & = &\displaystyle 
{\partial {\cal H}(q(x),p(x))\over \partial x^{\beta}}
- {\partial \over \partial x^{\alpha}}\left\langle
p(x),{\partial q(x)\over \partial x^1}\wedge \dots \wedge
{\partial q(x)\over \partial x^{\alpha -1}}\wedge
{\partial \over \partial x^{\beta}}\wedge
{\partial q(x)\over \partial x^{\alpha +1}}\wedge \dots \wedge
{\partial q(x)\over \partial x^n}\right\rangle \\
 & = & \displaystyle
{\partial {\cal H}(q(x),p(x))\over \partial x^{\beta}}
- \left\langle
{\partial p(x)\over \partial x^{\alpha}},
{\partial q(x)\over \partial x^1}\wedge \dots \wedge
{\partial q(x)\over \partial x^{\alpha -1}}\wedge
{\partial \over \partial x^{\beta}}\wedge
{\partial q(x)\over \partial x^{\alpha +1}}\wedge \dots \wedge
{\partial q(x)\over \partial x^n}\right\rangle .
\end{array}$$
Now assume that $u$ is a critical point, then because of (\ref{stress}) and (\ref{dH}),

$${\partial S^{\alpha}_{\beta}\over \partial x^{\alpha}}(x) =
{\partial L\over \partial x^{\beta}}(x,u(x),du(x)) =
- {\partial {\cal H}\over \partial x^{\beta}}(q(x),p(x)).$$
And we obtain

$$
\left\langle
{\partial p\over \partial x^{\alpha}},
{\partial q\over \partial x^1}\wedge \dots \wedge
{\partial q\over \partial x^{\alpha -1}}\wedge
{\partial \over \partial x^{\beta}}\wedge
{\partial q\over \partial x^{\alpha +1}}\wedge \dots \wedge
{\partial q\over \partial x^n}\right\rangle  
- {\partial \over \partial x^{\beta}}\left( {\cal H}(q,p)\right)
= - {\partial {\cal H}\over \partial x^{\beta}}(q,p)
$$
or equivalentely

\begin{equation}\label{hamilton4}
\sum_{\alpha}
\sum_{\footnotesize \begin{array}{c}\mu_1<\dots <\mu_n\\\mu_{\alpha} = \beta \end{array}}
{\partial (q^{\mu_1},\dots ,q^{\mu_{\alpha -1}},
p_{\mu_1\dots \mu_n}, q^{\mu_{\alpha +1}},\dots ,
q^{\mu_n})\over \partial (x^1,\dots ,x^n)} 
- {\partial \over \partial x^{\beta}}\left( {\cal H}(q,p)\right)
= - {\partial {\cal H}\over \partial x^{\beta}}(q,p).
\end{equation}

\noindent {\bf Conclusion} The Hamilton equations (\ref{hamilton3}) can be
completed by adding (\ref{hamilton4}) (which are actually a consequence of
(\ref{hamilton3})). We thus obtain 

\begin{equation}\label{hamilton5}
\begin{array}{c}\displaystyle
{\partial (q^{\mu_1},\dots ,q^{\mu_n})\over \partial (x^1,\dots ,x^n)} =
{\partial {\cal H}\over \partial p_{\mu_1\dots \mu_n}}(q,p)\\
\\
\displaystyle
\sum_{\alpha}
\sum_{\footnotesize \begin{array}{c}\mu_1<\dots <\mu_n\\\mu_{\alpha} = \nu\end{array}}
{\partial (q^{\mu_1},\dots ,q^{\mu_{\alpha -1}},
p_{\mu_1\dots \mu_n}, q^{\mu_{\alpha +1}},\dots ,
q^{\mu_n})\over \partial (x^1,\dots ,x^n)}
- \sum_{\alpha}\delta^{\alpha}_{\nu}
{\partial \over \partial x^{\alpha}}\left( {\cal H}(q,p)\right)
= - {\partial {\cal H}\over \partial q^{\nu}}(q,p).
\end{array}
\end{equation}

\subsection{The Cartan-Poincar\'e and pataplectic forms
on\\${\cal M} = \Lambda^nT^{\star}({\cal X}\times {\cal Y})$}
Motivated by the previous contruction, we define the Cartan-Poincar\'e form on
$\Lambda^nT^{\star}({\cal X}\times {\cal Y})$ to be

$$\theta := \sum_{\mu_1<\dots <\mu_n}p_{\mu_1\dots \mu_n}dq^{\mu_1}\wedge \dots \wedge dq^{\mu_n}.$$
Its differential is

$$\Omega := \sum_{\mu_1<\dots <\mu_n}dp_{\mu_1\dots \mu_n}\wedge dq^{\mu_1}\wedge \dots \wedge dq^{\mu_n},$$
which we will call the {\em pataplectic form}, a straightforward generalization
of the symplectic form.\\

A first property is that we can express the system of Hamilton's equations (\ref{hamilton5}) in an elegant way using $\Omega$. For any $(q,p)\in {\cal M}$ and any $n$-vector
$X\in \Lambda ^nT_{(q,p)}{\cal M}$
we define $X\inn \Omega \in T^{\star}_{(q,p)}{\cal M}$ as follows.
If $X$ is decomposable, i. e. if there exist $n$ vectors $X_1,\dots ,X_n\in T_{(q,p)}{\cal M}$ such that
$X = X_1\wedge \dots \wedge X_n$, we let

$$X\inn \Omega(V) := \Omega(X_1,\dots ,X_n,V),\quad \forall V\in T_{(q,p)}{\cal M}.$$
We extend this definition to non decomposable $X$ by linearity. Let us analyse $X\inn \Omega$ using
coordinates. Writing $X$ as 

{\small
$$\begin{array}{l}
\displaystyle \sum_{\mu_1<\dots <\mu_n}X^{\mu_1\dots \mu_n}
\partial _{\mu_1}\wedge \dots \wedge \partial _{\mu_n}\\
+\displaystyle \sum_{\footnotesize \begin{array}{c}\mu_1<\dots <\mu_{\alpha-1}<\mu_{\alpha+1}<\dots <\mu_n \\ \nu_1<\dots <\nu_n\end{array}}
X{^{\mu_1\dots \mu_{\alpha-1}}}{_{ \{\nu_1\dots \nu_n\} }} {^{\mu_{\alpha+1}\dots \mu_n}}
\partial _{\mu_1}\wedge \dots \wedge \partial _{\mu_{\alpha-1}}
\wedge \partial ^{\nu_1\dots \nu_n}\wedge \partial _{\mu_{\alpha+1}}
\wedge \dots \wedge \partial _{\mu_n}\\
+\quad \hbox{etc }\dots
\end{array}$$}
with the notations $\partial _{\mu}:= {\partial \over \partial q^{\mu}}$,
$\partial ^{\nu_1\dots \nu_n}:= {\partial \over \partial p_{\nu_1\dots \nu_n}}$, we have

$$X\inn\Omega = (-1)^n\left[
\sum_{\mu_1<\dots <\mu_n}X^{\mu_1\dots \mu_n} dp_{\mu_1\dots \mu_n}
- \sum_{\nu}\sum_{\alpha}
\sum_{\footnotesize \begin{array}{c}\mu_1<\dots <\mu_n\\\mu_{\alpha} = \nu\end{array}}
X{^{\mu_1\dots \mu_{\alpha-1}}} {_{\{\mu_1\dots \mu_n\} }} {^{\mu_{\alpha+1}\dots \mu_n}} dq^{\nu}
\right] .$$
Algebraic similarities with (\ref{hamilton5}) are evident if we replace $X$ by
${\partial (q,p)\over \partial (x^1,\dots ,x^n)}:=
{\partial (q,p)\over \partial x^1}\wedge \dots
\wedge {\partial (q,p)\over \partial x^n}$. In particular we can see easily that the coefficients of $dy^i$ and $dp_{\mu_1\dots \mu_n}$ in
$(-1)^n{\partial (q,p)\over \partial (x^1,\dots ,x^n)}\inn \Omega$
and $d{\cal H}$ coincide if and only if the Hamilton system (\ref{hamilton3}) holds.
Thus we are led to define ${\cal I}$ to be the algebraic ideal in
$\Lambda^{\star}{\cal M}$ spanned by $\{dx^1,\dots ,dx^n\}$ and hence (\ref{hamilton3})
is equivalent to

\begin{equation}\label{hamilton7}
(-1)^n{\partial (q,p)\over \partial (x^1,\dots ,x^n)}\inn \Omega =
d{\cal H} \quad \hbox{mod }{\cal I}.
\end{equation}

\begin{defi}
A $n$-vector $X\in \Lambda^nT_{(q,p)}{\cal M}$ is ${\cal H}$-Hamiltonian if
and only if

\begin{equation}\label{hamilvec1}
(-1)^nX\inn \Omega = d{\cal H}\quad \hbox{mod }{\cal I}.
\end{equation}
\end{defi}
For such an $X$, it is possible to precise the relation between the left and right hand sides of
(\ref{hamilvec1}) in the case where $X$ is decomposable, i. e.
$X = X_1\wedge \dots \wedge X_n$. Notice that (\ref{hamilvec1})
implies in particular $X^{1\dots n} = {\partial {\cal H}\over \partial \epsilon} = 1$ (where $ \epsilon := p_{1\dots n}$ see (\ref{hamilton5})),
which is equivalent to $\omega(X_1,\dots ,X_n)=1$. Hence we may
always assume without loss of generality that the $X_{\alpha}$ are chosen
so that $dx^{\beta}(X_{\alpha}) = \delta^{\beta}_{\alpha}$. Such vectors are unique.

\begin{lemm}
Let $X = X_1\wedge \dots \wedge X_n\in \Lambda^nT_{(q,p)}{\cal M}$ such that
$dx^{\beta}(X_{\alpha}) = \delta^{\beta}_{\alpha}$. Assume that $X$ is
${\cal H}$-Hamiltonian, then

\begin{equation}\label{hamilvec2}
(-1)^nX\inn \Omega = d{\cal H} - \sum_{\alpha}d{\cal H}(X_{\alpha})dx^{\alpha}.
\end{equation}
\end{lemm}
{\bf Proof} Since for any $\alpha$, $\beta$,
$dx^{\beta}\left( X_{\alpha} - {\partial \over \partial x^{\alpha}}\right) =0$,
equation (\ref{hamilvec1}) implies that for all $\alpha$,

$$(-1)^nX\inn \Omega\left( X_{\alpha} - {\partial \over \partial x^{\alpha}}\right) =
d{\cal H}\left( X_{\alpha} - {\partial \over \partial x^{\alpha}}\right)$$
$$\Longleftrightarrow
(-1)^n\Omega\left( X_1,\dots ,X_n,X_{\alpha} - {\partial \over \partial x^{\alpha}}\right)
= d{\cal H}(X_{\alpha}) - {\partial {\cal H}\over \partial x^{\alpha}}$$
$$\Longleftrightarrow
(-1)^nX\inn \Omega\left( {\partial \over \partial x^{\alpha}}\right) =
{\partial {\cal H}\over \partial x^{\alpha}} - d{\cal H}(X_{\alpha}).$$
This implies

\begin{equation}\label{hamilvec0}
(-1)^n\sum_{\alpha}X\inn \Omega
\left( {\partial \over \partial  x^{\alpha}}\right) dx^{\alpha} = 
\sum_{\alpha}{\partial {\cal H}\over \partial x^{\alpha}}dx^{\alpha}
- \sum_{\alpha}d{\cal H}(X_{\alpha})dx^{\alpha}.
\end{equation}
Now if we rewrite (\ref{hamilvec1}) as

\noindent $\displaystyle (-1)^n\left( \sum_iX\inn \Omega
\left( {\partial \over \partial  y^i}\right) dy^i +
\sum_{\mu_1<\dots  <\mu_n}X\inn \Omega
\left( {\partial \over \partial  p_{\mu_1\dots  \mu_n}}\right) dp_{\mu_1\dots  \mu_n}
\right) =$
$$\sum_i {\partial {\cal H}\over \partial y^i}dy^i +
\sum_{\mu_1<\dots  <\mu_n} {\partial {\cal H}\over \partial p_{\mu_1\dots  \mu_n}}
dp_{\mu_1\dots  \mu_n},$$
and sum with (\ref{hamilvec0}), we obtain exactly (\ref{hamilvec2}).
\bbox

As a Corollary of this result we deduce that a reformulation of
(\ref{hamilton7}) is

\begin{equation}\label{hamilton6}
(-1)^n{\partial (q,p)\over \partial (x^1,\dots ,x^n)}\inn \Omega =
d{\cal H} - \sum_{\alpha}d{\cal H}
\left( {\partial (q,p)\over \partial x^{\alpha}}\right) dx^{\alpha}.
\end{equation}
It is an exercise to check that actually this relation is a direct translation
of (\ref{hamilton5}).\\

\subsection{A variational formulation of (\ref{hamilton5})}

We shall now prove that equations (\ref{hamilton5}) are the Euler-Lagrange equations of some
simple functional. For that purpose, let $\Gamma$ be an oriented submanifold of dimension $n$
in $\Lambda^nT^{\star}({\cal X}\times {\cal Y})$ such that $\omega_{|\Gamma}>0$ everywhere.
Then we define the functional

$${\cal A}[\Gamma] := \int_{\Gamma}\theta -\lambda{\cal H}(q,p)\omega.$$
Here $\lambda$ is a (real) scalar function defined over $\Gamma$ which plays
the role of a Lagrange multiplier. We now characterise submanifolds $\Gamma$ which are critical points
of ${\cal A}$.\\

\noindent {\bf Variations with respect to $p$}\\
Let $\delta p$ be some infinitesimal variation of $\Gamma$ with compact support. We compute

$$\delta{\cal A}_{\Gamma}(\delta p) = \int_{\Gamma}\delta p_{\mu_1\dots \mu_n}
\left( dq^{\mu_1}\wedge \dots \wedge dq^{\mu_n}
-\lambda {\partial {\cal H}\over \partial p_{\mu_1\dots \mu_n}}\omega \right) .$$
Assuming that this vanishes for all $\delta p$, we obtain

$$\left( dq^{\mu_1}\wedge \dots \wedge dq^{\mu_n}\right) _{|\Gamma} =
\lambda {\partial {\cal H}\over \partial p_{\mu_1\dots \mu_n}}\omega_{|\Gamma}.$$
This relation means that for any orientation preserving parametrization
$(t^1,\dots ,t^n)\longmapsto (q,p)(t^1,\dots ,t^n)$
of $\Gamma$,

$${\partial (q^{\mu_1},\dots ,q^{\mu_n})\over \partial (t^1,\dots ,t^n)}
= \lambda {\partial {\cal H}\over \partial p_{\mu_1\dots \mu_n}}\omega
\left( {\partial q^{\mu_1}\over \partial t^1},\dots ,{\partial q^{\mu_n}\over \partial t^n}\right) .$$
But we remark that because ${\partial {\cal H}\over \partial p_{1\dots n}} = 1$, the above relation for
$(\mu_1,\dots ,\mu_n) = (1,\dots ,n)$ forces $\lambda =1$. Hence

$${\cal A}[\Gamma] = \int _{\Gamma}\theta -{\cal H}(q,p)\omega.$$
Moreover the equation obtained here can be written using the natural parametrization
$(x^1,\dots ,x^n)\longmapsto (x,u(x),p(x))$ (for which
$\omega\left( {\partial \over \partial x^1},\dots ,{\partial \over \partial x^n}\right) =1$)
and then we obtain

$${\partial q\over \partial x^1}\wedge \dots \wedge {\partial q\over \partial x^n} =
{\partial {\cal H}\over \partial p}(q,p),$$
i. e. exactly equation (\ref{hamilton1})
\footnote{Note that this relation actually implies
${\cal A}[\Gamma] = \int_{\cal X}L(x,q,dq)\omega$. Hence, as in the
one-dimensional Hamilton formalism, $\theta - {\cal H}\omega$ plays the role of
the Lagrangian density.}.\\

\noindent {\bf Variations with respect to $q$}\\
Now $\delta q$ is some infinitesimal variation of $\Gamma$ with compact support. And we have

$$\delta{\cal A}_{\Gamma}(\delta q) = \int_{\Gamma}\sum_{\mu_1<\dots <\mu_n}\sum_{\alpha}
p_{\mu_1\dots \mu_n}dq^{\mu_1}\wedge \dots \wedge d(\delta q^{\mu_{\alpha}})\wedge \dots  \wedge dq^{\mu_n}
- \sum_{\mu}{\partial {\cal H}\over \partial q^{\mu}}\delta q^{\mu}\omega - {\cal H}(q,p)\delta\omega.$$
We pay  special attention to $\delta\omega$:

$$\delta\omega = d(\delta x^1)\wedge \dots \wedge dx^n + \dots + dx^1\wedge \dots \wedge d(\delta x^n).$$
Hence

$$\begin{array}{ccl}
\displaystyle \int_{\Gamma}{\cal H}(q,p)\delta\omega & = & \displaystyle - \int_{\Gamma}
\delta x^1\left( d({\cal H}(q,p))\wedge \dots \wedge dx^n\right) + \dots +
\delta x^n\left( dx^1\wedge \dots \wedge d({\cal H}(q,p))\right) \\
& = & \displaystyle - \sum _{\alpha}\delta x^{\alpha}{\partial \over \partial x^{\alpha}}\left( {\cal H}(q,p)\right) \omega.
\end{array}$$
Thus after integrations by parts, we obtain

$$\begin{array}{ccl}
\displaystyle \delta{\cal A}_{\Gamma}(\delta q) & = & \displaystyle
\int_{\Gamma} - \sum_{\mu_1<\dots <\mu_n}\sum_{\alpha}\delta q^{\mu_{\alpha}}
dq^{\mu_1}\wedge \dots \wedge dq^{\mu_{\alpha -1}}\wedge dp_{\mu_1\dots \mu_n}\wedge dq^{\mu_{\alpha +1}}
\wedge \dots  \wedge dq^{\mu_n}\\
& & \displaystyle - \sum_{\mu}{\partial {\cal H}\over \partial q^{\mu}}\delta q^{\mu}\omega
+\sum _{\alpha}\delta x^{\alpha}{\partial \over \partial x^{\alpha}}\left( {\cal H}(q,p)\right) \omega.
\end{array}$$
And this vanishes if and only if

$$\sum_{\alpha}\sum_{\begin{array}{c}\mu_1<\dots <\mu_n\\\mu_{\alpha} = \nu\end{array}}
dq^{\mu_1}\wedge \dots \wedge dq^{\mu_{\alpha -1}}\wedge dp_{\mu_1\dots \mu_n}\wedge dq^{\mu_{\alpha +1}}
\wedge \dots  \wedge dq^{\mu_n} -
\sum _{\alpha}\delta^{\alpha}_{\nu}{\partial \over \partial x^{\alpha}}\left( {\cal H}(q,p)\right) \omega=
- {\partial {\cal H}\over \partial q^{\nu}}\omega.$$
Again by choosing the parametrization $(x^1,\dots ,x^n)\longmapsto (x,u(x),p(x))$, this relation is
easily seen to be equivalent to (\ref{hamilton2}) and (\ref{hamilton4}).

By the same token we have proven that if we look to critical points of the functional
$\int_{\Gamma}\theta$ with the constraint ${\cal H}(q,p) = h$, for some constant $h$, then the Lagrange
multiplier is 1 and they satisfy the same equations.

\begin{theo}
Let $\Gamma$ be an oriented submanifold of dimension $n$
in $\Lambda^nT^{\star}({\cal X}\times {\cal Y})$ such that $\Omega_{|\Gamma}>0$ everywhere.
Then the three following assertions are equivalent
\begin{itemize}
\item $\Gamma$ is the graph of a solution of the generalized Hamilton equations (\ref{hamilton5})
\item $\Gamma$ is a critical point of the functional
$\int_{\Gamma}\theta - {\cal H}(q,p)\omega$
\item $\Gamma$ is a critical point of the functional $\int_{\Gamma}\theta$ under the constraint
that ${\cal H}(q,p)$ is constant.
\end{itemize}
\end{theo}

\subsection{Some particular cases}
By restricting the variables $(q,p)$ to lie in some submanifold of
${\cal M}=\Lambda^nT^{\star}({\cal X}\times {\cal Y})$,
the Legendre correspondance becomes in some situations a true map.\\

\noindent {\bf a)} We assume that all components $p_{\mu_1,\dots \mu_n}$ vanishes excepted for

$$p_{1\dots n} =: \epsilon \quad \hbox{and}\quad p_{1\dots (\alpha-1)(n+i)(\alpha+1)\dots n} =: p^{\alpha}_i$$
and all obvious permutations in the indices. This defines a submanifold ${\cal M}_{\hbox {\tiny Weyl}}$ of
${\cal M}$. It means that

$$\theta_{|{\cal M}_{\hbox {\tiny Weyl}}}= \epsilon\; dx^1\wedge \dots \wedge dx^n +
\sum_{\alpha}\sum_ip^{\alpha}_idx^1\wedge \dots \wedge dx^{\alpha-1}\wedge dy^i
\wedge dx^{\alpha+1}\wedge \dots \wedge dx^n.$$
Then for any $(q,p)\in {\cal M}_{\hbox {\tiny Weyl}}$,
$\langle p,z_1\wedge \dots \wedge z_n\rangle = \epsilon + \sum_{\alpha}\sum_ip^{\alpha}_iv^i_{\alpha}$,
$W(q,v,p) = \epsilon + \sum_{\alpha}\sum_ip^{\alpha}_iv^i_{\alpha} - L(q,v)$. Hence the relation
(\ref{dlp}) ${\partial W\over \partial v^i_{\alpha}}(q,v)=0$ is equivalent to

$$p^{\alpha}_i = {\partial L\over \partial v^i_{\alpha}}(q,v)
\Longleftrightarrow v^i_{\alpha} = {\cal V}^i_{\alpha}(q,p) .$$
The relation (\ref{lw}) $W(q,v,p) = w$ gives

$$\epsilon = w + L(q,v) - \sum_{\alpha}\sum_i{\partial L\over \partial v^i_{\alpha}}(q,v)v^i_{\alpha}.$$
Last we have that
${\cal H}(q,p) = \epsilon + \sum_{\alpha}\sum_ip^{\alpha}_i{\cal V}^i_{\alpha}(q,p) - L(q,{\cal V}(q,p))$.

This example shows that  for any
$(q,v,w)\in S\Lambda^nT({\cal X}\times {\cal Y})\times \Bbb{R}$,
there exist $(q,p)\in {\cal M}$ such that $(q,v,w)\longleftrightarrow (q,p)$ and
this $(q,p)$ is unique
if it is chosen in ${\cal M}_{\hbox {\tiny Weyl}}$.\\

To summarize, we recover the Weyl theory (see \cite{Rund,Helein}).
As an exercize, the reader can check that
in this situation, equations (\ref{hamilton3}) are equivalent to

\begin{equation}\label{weyl}
{\partial y^i\over \partial x^{\alpha}} = {\partial {\cal H}\over \partial p^{\alpha}_i},\quad
\sum_{\alpha}{\partial p^{\alpha}_i\over \partial x^{\alpha}} = - {\partial {\cal H}\over \partial y^i}.
\end{equation}

\noindent {\bf b)}  We assume that $(q,p)$ are such that there exist coefficients $\left( \pi^{\alpha}_{\mu}\right)_{\alpha,\mu}$
with

$$p_{\mu_1\dots \mu_n} = \left| \begin{array}{ccc}
\pi^1_{\mu_1} & \dots & \pi^1_{\mu_n}\\
\vdots & & \vdots \\
\pi^n_{\mu_1} & \dots & \pi^n_{\mu_n}\\
\end{array}\right| .$$
This constraint defines a submanifold ${\cal M}_{\hbox {\tiny Carath\'eodory}}$ of ${\cal M}$. Then

$$\theta_{|{\cal M}_{\hbox {\tiny Carath\'eodory}}} = \left( \sum_{\mu_1}\pi^1_{\mu_1}dq^{\mu_1}\right) \wedge \dots \wedge
\left( \sum_{\mu_n}\pi^1_{\mu_n}dq^{\mu_n}\right) .$$
Then it is an exercise to see that, by choosing $w=0$, it leads
to the formalism developped in \cite{Rund} and \cite{Helein}
associated to the Carath\'eodory
theory of equivalent integrals. However it is not clear in general whether it is
possible to perform the Legendre transform in this setting by being able
to fix arbitrarirely the value of $w$. It is so if we do not impose a 
condition on $w$.

\section{Comparison with the usual Hamiltonian formalism for quantum fields theory}
\subsection{Reminder of the usual approach to quantum field theory}
Here we compare the preceeding construction with the classical approach to quantum field theory
by  so-called canonical quantization. We shall first explore it in the case where ${\cal X}$ is the
Minkowski space $\Bbb{R}\times \Bbb{R}^{n-1}$ and $y=\phi$ is a real scalar field. Hence ${\cal Y}=\Bbb{R}$.
Our functional is

$${\cal L}[\phi] := \int_{\Bbb{R}\times \Bbb{R}^{n-1}}L(x,\phi,d\phi)dx.$$
For simplicity, we may keep in mind the following example of Lagrangian:

$$\int_{\Bbb{R}\times \Bbb{R}^{n-1}}L(x,\phi,d\phi)dx =
\int_{\Bbb{R}\times \Bbb{R}^{n-1}}\left( {1\over 2}\left( {\partial \phi\over \partial x^0}\right) ^2
- {1\over 2}\sum_{\alpha=1}^{n-1}\left( {\partial \phi\over \partial x^{\alpha}}\right) ^2 - V(\phi)\right) dx^0d\vec{x},$$
where we denote $\vec{x} = (x^{\alpha})_{1\leq \alpha\leq n-1}$. We shall also denote $t=x^0$.

The usual approach consists in selecting a global time coordinate $t$ as we already
implicitely assumed here. Then for each time the instantaneous state of the field is
seen as a point in the infinite dimensional ``manifold''
$\mathfrak{F}:= \{\Phi:\Bbb{R}^{n-1}\longrightarrow \Bbb{R}\}$.
Hence we view the field $\phi$ rather as a path

$$\begin{array}{ccl}
\Bbb{R} & \longrightarrow & \mathfrak{F}\\
t & \longmapsto & [\vec{x}\longmapsto \phi(t,\vec{x}) = \Phi^{\vec{x}}(t)].
\end{array}$$
We thus recover the problem of studying the dynamics of a point moving in a configuration space $\mathfrak{F}$. The prices to
pay are 1) $\mathfrak{F}$ is infinite dimensional 2) we lose  relativistic invariance.\\

In this viewpoint, ${\cal L}[\phi] =
\int_{\Bbb{R}}\mathfrak{L}[t,\Phi(t),{d\Phi\over dt}(t)]dt$, where
$\Phi(t) = [\vec{x}\longmapsto \phi(t,\vec{x})]\in \mathfrak{F}$,
${d\Phi\over dt}(t) = [\vec{x}\longmapsto {\partial \phi\over \partial t}(t,\vec{x})]\in T_{\Phi(t)}\mathfrak{F}$ and
$\mathfrak{L}[t,\Phi(t),{d\Phi\over dt}(t)] = \int_{\Bbb{R}^{n-1}}L(x,\phi(x),d\phi(x))d\vec{x}$.\\

Then we consider the ``symplectic'' manifold which is formally $T^{\star}\mathfrak{F}$, i. e. we
introduce the dual variable

$$\Pi := {\partial \mathfrak{L}\over \partial {d\Phi\over dt}},$$
or equivalentely $\Pi(t) = [\vec{x}\longmapsto \pi(t,\vec{x})=\Pi_{\vec{x}}(t)]$ with

$$\Pi_{\vec{x}}(t) = {\partial \mathfrak{L}\over \partial {d\Phi^{\vec{x}}\over dt}}[t,\Phi(t),{d\Phi\over dt}(t)]\quad
\Longleftrightarrow \quad
\pi(t,\vec{x})={\delta \mathfrak{L}\over \delta {\partial \phi(t,\vec{x})\over \partial t}}[t,\Phi(t),{d\Phi\over dt}(t)]
= {\partial L\over \partial v_0}(x,\phi(x),d\phi(x)).$$
Here ${\delta \over \delta \phi(\vec{x})}$ is the Fr\'echet derivative. In our example

$$\Pi_{\vec{x}}(t) = {\partial \phi\over \partial t}(t,\vec{x}).$$

We define the Hamiltonian functional to be

$$\mathfrak{H}[\Phi,\Pi] := \int_{\Bbb{R}^{n-1}}\Pi_{\vec{x}}{\dot{\Phi}}^{\vec{x}}d\vec{x} - \mathfrak{L}[t,\Phi, {d\Phi\over dt}]
= \int_{\Bbb{R}^{n-1}}\left( {1\over 2}\pi(\vec{x})^2+{1\over 2}|\nabla \phi (\vec{x})|^2 + V(\phi(\vec{x}))\right)  d\vec{x}.$$

Now we can write the equations of motion as

$$\begin{array}{ccccl}
\displaystyle {\partial \pi\over \partial t}(t,\vec{x}) = {d\Pi_{\vec{x}}\over dt} & = &\displaystyle 
- {\partial \mathfrak{H}\over \partial \Phi^{\vec{x}}}(\Phi,\Pi)  & = & \Delta \phi - V'(\phi)\\
\displaystyle {\partial \phi\over \partial t}(t,\vec{x}) = {d\Phi^{\vec{x}}\over dt} & = &
\displaystyle {\partial \mathfrak{H}\over \partial \Pi^{\vec{x}}}(\Phi,\Pi) & = & \pi(t,\vec{x}).
\end{array}$$
A Poisson bracket can be defined on the set of functionals $\{A:T^{\star}\mathfrak{F}\longmapsto \Bbb{R}\}$
by

$$\{A,B\} :=  \int_{\Bbb{R}^{n-1}}\left( {\delta A\over \delta \pi(\vec{x})}{\delta B\over \delta \phi(\vec{x})}
- {\delta A\over \delta \phi(\vec{x})}{\delta B\over \delta \pi(\vec{x})}\right) d\vec{x},$$
where ${\delta A\over \delta \phi(\vec{x})}$ is the Fr\'echet derivative with respect to $\phi(\vec{x})$, i. e. 
the distribution such that for any smooth compactly supported deformation $\delta \phi$ of $\phi$,

$$dA_{\phi}[\delta \phi] = \int_{\Bbb{R}^{n-1}}\delta \phi(\vec{x}){\delta A\over \delta \phi(\vec{x})}d\vec{x}.$$

And we may formulate the dynamical equations using the Poisson bracket as

$$\begin{array}{ccl}
\displaystyle {d\Pi_{\vec{x}}\over dt} & = & \displaystyle \{\mathfrak{H},\Pi_{\vec{x}}\}\\
\displaystyle {d\Phi^{\vec{x}}\over dt} & = &\displaystyle \{\mathfrak{H},\Phi^{\vec{x}}\},
\end{array}$$
with

$$\{\Phi^{\vec{x}},\Phi^{\vec{x}'}\} = \{\Pi_{\vec{x}},\Pi_{\vec{x}'}\} = 0,\quad
\{\Pi_{\vec{x}},\Phi^{\vec{x}'}\} = \delta_{\vec{x}}^{\vec{x}'} = \delta^{n-1}(\vec{x} - \vec{x}').$$
This singular Poisson bracket means that for any test functions $f,g\in {\cal C}^{\infty}_c(\Bbb{R}^{n-1},\Bbb{R})$,

$$\left\{ \int_{\Bbb{R}^{n-1}}g(\vec{x})\Pi_{\vec{x}}d\vec{x},
\int_{\Bbb{R}^{n-1}}f(\vec{x}')\Phi^{\vec{x}'}d\vec{x}'
\right\} = 
\int_{\Bbb{R}^{n-1}} f(\vec{x})g(\vec{x})d\vec{x}.$$
This implies in particular

$$\left\{ \int_{\Bbb{R}^{n-1}}g(\vec{x})\Pi_{\vec{x}}d\vec{x},
\int_{\Bbb{R}^{n-1}}f(\vec{x}')V(\Phi^{\vec{x}'})d\vec{x}'
\right\} = 
\int_{\Bbb{R}^{n-1}} V'(\Phi^{\vec{x}})f(\vec{x})g(\vec{x})d\vec{x},$$
because of the derivation property of the Poisson bracket.\\

\subsection{Translation in pataplectic geometry}

We first adapt and modify our notations: the coordinates on
${\cal M}=\Lambda^nT^{\star}(\Bbb{R}\times \Bbb{R}^{n-1}\times \Bbb{R})$
are now written $(q^{\mu},p_{\mu_1\dots \mu_n}) =  (x^{\alpha}, y,\epsilon,p^{\alpha} )$
where $0\leq \alpha \leq n-1$, $q^0=x^0=t$,
$(x^{\alpha})_{1\leq \alpha\leq n-1}=\vec{x}$, $q^n=y$ and

$$\epsilon:= p_{0\dots (n-1)}\quad p^{\alpha}:= p_{0\dots (\alpha-1)n(\alpha+1)\dots (n-1)}.$$
Hence

$$\theta = \epsilon\; dx^0\wedge \dots \wedge dx^{n-1} + \sum_{\alpha=0}^{n-1}
p^{\alpha}dx^0\wedge \dots \wedge dx^{\alpha-1}\wedge dy\wedge dx^{\alpha+1}\wedge \dots \wedge dx^{n-1},$$
or letting $\omega:= dx^0\wedge \dots \wedge dx^{n-1}$ and
$\omega_{\alpha}:= (-1)^{\alpha}dx^0\wedge \dots \wedge dx^{\alpha-1}\wedge dx^{\alpha+1}\wedge \dots \wedge dx^{n-1}
={\partial \over \partial x^{\alpha}}\inn \omega$,

$$\theta = \epsilon\; \omega + \sum_{\alpha=0}^{n-1}p^{\alpha}dy\wedge \omega_{\alpha}\quad
\hbox{and}\quad
\Omega = d\epsilon \wedge \omega + \sum_{\alpha=0}^{n-1}dp^{\alpha}\wedge dy\wedge \omega_{\alpha}.$$

Thus we see that in the present case the pataplectic formalism reduces essentially to the Weyl
formalism, because the fields are one dimensional.

Let us consider some field $\phi$, a map \footnote{In most applications
it will be more suitable to assume $w=0$, but it is useful here
to keep $w$ arbitrary for the moment.} 
$w:\Bbb{R}\times \Bbb{R}^{n-1}\longrightarrow \Bbb{R}$ and
$p$ such that $(x,\phi(x),d\phi(x))\leftrightarrow (x,\phi(x),p(x))$. This means that we are forced to have

$$p^{\alpha} = {\partial L\over \partial v_{\alpha}}(x,\phi(x),d\phi(x))\quad \hbox{and} \quad
\epsilon = w +L(x,\phi(x),d\phi(x)) - \sum_{\alpha=0}^{n-1}p^{\alpha}{\partial \phi\over \partial x^{\alpha}}(x).$$

We let $\Gamma:= \{(x,\phi(x),p(x))/x\in \Bbb{R}\times \Bbb{R}^{n-1}\}
\subset {\cal M}$ and we consider the instantaneous
slices $S_t:= \Gamma \cap \{x^0=t\}$. These slices are oriented by the condition
${\partial \over \partial t}\inn \omega_{|S_t}>0$.
Then we can express the observables

$$\Phi^f(t):= \int_{\Bbb{R}^{n-1}}f(\vec{x})\Phi^{\vec{x}}(t)d\vec{x},\quad 
\Pi_g(t):= \int_{\Bbb{R}^{n-1}}g(\vec{x})\Pi_{\vec{x}}(t)d\vec{x}$$
and

$$\mathfrak{H}[\Phi(t),\Pi(t)] = \int_{\Bbb{R}^{n-1}}\left( \pi(t,\vec{x}){\partial \phi\over \partial t}(t,\vec{x})
-L(t,\vec{x},\phi(x),d\phi(x))\right) d\vec{x} =
\int_{\Bbb{R}^{n-1}}H^0_0(t,\vec{x},\phi)\omega_0$$
as integrals of $(n-1)$-forms on $S_t$. First

$$\Phi^f(t) = \int_{S_t}f(\vec{x})\phi(t,\vec{x})dx^1\wedge \dots \wedge dx^{n-1}
 = \int_{S_t}Q^f,\quad \hbox{with }Q^f:= f(\vec{x})\; y\; \omega_0.$$

$$\Pi_g(t) = \int_{S_t}g(\vec{x})\pi(t,\vec{x})dx^1\wedge \dots \wedge dx^{n-1} =
\int_{S_t}P_g,
\quad \hbox{with }P_g:= g(\vec{x})\sum_{\alpha=0}^{n-1}p^{\alpha}\omega_{\alpha},$$
because $\pi(t,\vec{x}) = {\partial L\over \partial v_0}(x,\phi(x),d\phi(x)) = p^0$
and $\omega_{\alpha|S_t} = 0$ if $\alpha\geq 1$\\

\noindent And last

$$\mathfrak{H}[\Phi(t),\Pi(t)] = \int_{\Bbb{R}^{n-1}}{\cal H}(q,p)\omega_0 -
\int_{\Bbb{R}^{n-1}}\epsilon \omega_0 +\sum_{\alpha=1}^{n-1}p^{\alpha}dy\wedge
\left( {\partial \over \partial x^{\alpha}}\inn \omega_0\right) =
\int_{S_t}\eta_0,$$
where

$$\eta_0:= {\cal H}(q,p)\omega_0 -\left( \epsilon \omega_0 +\sum_{\alpha=1}^{n-1}p^{\alpha}
dx^1\wedge \dots \wedge dx^{\alpha-1}\wedge dy\wedge dx^{\alpha+1}\wedge \dots \wedge dx^{n-1}\right) ,$$
because $H^0_0(x,\phi) = {\cal H}(q,p) -\langle p,{\partial \over \partial t}\wedge z_1\wedge \dots \wedge z_{n-1}\rangle
={\cal H}(q,p) - \left( \epsilon +\sum_{\alpha=1}^{n-1}p^{\alpha}{\partial \phi\over \partial x^{\alpha}}\right)$.

We remark \footnote{we observe also that
$P_g = g(\vec{x}){\partial \over \partial y}\inn \theta =
g(\vec{x}){\partial \over \partial y}\inn
(\theta - {\cal H}(q,p)\omega)$.} that

$$\eta_0 = - {\partial \over \partial t}\inn (\theta - {\cal H}(q,p)\omega).$$

\subsection{Recovering  the usual Poisson brackets as a local expression}

\noindent Our aim is now to express the various Poisson brackets involving
the quantities $\Phi^f(t)$ and $\Pi_g(t)$ along $\Gamma$ using some analogue of
the Poisson bracket defined on $(n-1)$-forms.
We generalize slightly the definition of $Q^f$ to be

\begin{equation}\label{3.3.f}
Q^f = \sum_{\alpha=0}^{n-1}f^{\alpha}(x)\; y\; \omega_{\alpha},
\end{equation}
where $f:= \sum_{\alpha=0}^{n-1}f^{\alpha}(x){\partial \over \partial x^{\alpha}}$
is some vector field. Hence our observables become

\begin{equation}\label{3.3.QP}
\Phi^f(t) = \int_{S_t}Q^f\quad \hbox{and}\quad \Pi_g(t) = \int_{S_t}P_g,
\end{equation}
where $P_g:= g(x)\sum_{\alpha=0}^{n-1}p^{\alpha}\omega_{\alpha}$ as before
\footnote{notice that
actually $\int_{S_t}Q^f = \int_{S_t}f^0(x)\; y\; \omega_0$.}.
We shall see here that we can define a bracket operation $\{.,.\}$ between
$Q^f$, $P_g$ and $\eta_0$ such that the usual Poisson bracket of fields
actually derives from $\{.,.\}$ by 

\begin{equation}\label{trans}
\int_{S_t}\{P_g,Q^f\} = \left\{ \int_{S_t}P_g,\int_{S_t}Q^f\right\},\;\hbox{etc}\dots
\end{equation}

\noindent First we remark that

$$\begin{array}{ccl}
dQ^f & = & \displaystyle \sum_{\alpha=0}^{n-1}f^{\alpha}\; dy\wedge \omega_{\alpha} +
\sum_{\alpha=0}^{n-1}y{\partial f^{\alpha}\over \partial x^{\alpha}}\omega 
= \sum_{\alpha=0}^{n-1}f^{\alpha}{\partial \over \partial p^{\alpha}}\inn \Omega +
\sum_{\alpha=0}^{n-1}y{\partial f^{\alpha}\over \partial x^{\alpha}}{\partial \over \partial \epsilon}\inn \Omega \\
& = & - \xi_{Q^f}\inn \Omega
\end{array}$$
and

$$\begin{array}{ccl}
dP_g & = & \displaystyle \sum_{\alpha=0}^{n-1}p^{\alpha}{\partial g\over \partial x^{\alpha}} \omega +
\sum_{\alpha=0}^{n-1}gdp^{\alpha}\wedge \omega_{\alpha} 
= \sum_{\alpha=0}^{n-1}p^{\alpha}{\partial g\over \partial x^{\alpha}} 
{\partial \over \partial \epsilon}\inn \Omega 
- g{\partial \over \partial y}\inn \Omega \\
& = & - \xi_{P_g}\inn \Omega,
\end{array}$$
where

\begin{equation}\label{3.3.A}
\xi_{Q^f}:= -\sum_{\alpha=0}^{n-1}f^{\alpha}{\partial \over \partial p^{\alpha}}
- y\sum_{\alpha=0}^{n-1}{\partial f^{\alpha}\over \partial x^{\alpha}}{\partial \over \partial \epsilon}
\end{equation}
and

\begin{equation}\label{3.3.B}
\xi_{P_g}:= g{\partial \over \partial y} -
\sum_{\alpha=0}^{n-1} p^{\alpha}{\partial g\over \partial x^{\alpha}}{\partial \over \partial \epsilon}.
\end{equation}
Also notice that

$$d\eta_0 = (d{\cal H}-d\epsilon)\wedge \omega_0 - \sum_{\alpha=1}^{n-1}dp^{\alpha}\wedge dy\wedge
\left( {\partial \over \partial x^{\alpha}}\inn \omega_0\right) .$$

\begin{defi} We define the Poisson $\mathfrak{p}$-brackets of these $(n-1)$-forms to be

$$\{\eta_0,Q^f\} := - \xi_{Q^f}\inn d\eta_0,\quad \{\eta_0,P_g\} := - \xi_{P_g}\inn d\eta_0,$$

$$\{P_g,Q^f\} := - \xi_{Q^f}\inn dP_g =  \xi_{P_g}\inn dQ^f  = \xi_{Q^f}\inn (\xi_{P_g}\inn \Omega)$$
and
$$\{Q^f,Q^{f'}\} := \xi_{Q^{f'}}\inn (\xi_{Q^f}\inn \Omega),\quad 
\{P_g,P_{g'}\} := \xi_{P_{g'}}\inn (\xi_{P_g}\inn \Omega).$$
\end{defi}

\noindent Let us now compute these $\mathfrak{p}$-brackets. We use in particular
the fact that ${\partial {\cal H}\over \partial \epsilon} = 1$.

$$\begin{array}{ccl}
\{\eta_0,Q^f\} & = & \displaystyle \left( \sum_{\alpha=0}^{n-1}f^{\alpha}{\partial \over \partial p^{\alpha}}
+ y\sum_{\alpha=0}^{n-1}{\partial f^{\alpha}\over \partial x^{\alpha}}{\partial \over \partial \epsilon}\right)
\inn \left( (d{\cal H}-d\epsilon)\wedge \omega_0 - \sum_{\alpha=1}^{n-1}dp^{\alpha}\wedge dy\wedge \left( {\partial \over \partial x^{\alpha}}\inn \omega_0\right) \right) \\
& = & \displaystyle \sum_{\alpha=0}^{n-1}f^{\alpha}{\partial {\cal H}\over \partial p^{\alpha}}\omega_0
- \sum_{\alpha=1}^{n-1}f^{\alpha}dy\wedge 
\left( {\partial \over \partial x^{\alpha}}\inn \omega_0\right) .
\end{array}$$

$$\begin{array}{ccl}
\{\eta_0,P_g\} & = & \displaystyle \left(
\sum_{\alpha=0}^{n-1} p^{\alpha}{\partial g\over \partial x^{\alpha}}{\partial \over \partial \epsilon}
- g{\partial \over \partial y}\right) 
\inn \left( (d{\cal H}-d\epsilon)\wedge \omega_0 - \sum_{\alpha=1}^{n-1}dp^{\alpha}\wedge dy\wedge \left( {\partial \over \partial x^{\alpha}}\inn \omega_0\right) \right) \\
& = & \displaystyle -g{\partial {\cal H}\over \partial y}\omega_0 - g\sum_{\alpha=1}^{n-1}dp^{\alpha}\wedge \left( {\partial \over \partial x^{\alpha}}\inn \omega_0\right) ,
\end{array}$$

$$\begin{array}{ccl}
\{P_g,Q^f\} & = & \displaystyle \left(
\sum_{\alpha=0}^{n-1}f^{\alpha}{\partial \over \partial p_{\alpha}} +
y\sum_{\alpha=0}^{n-1}{\partial f^{\alpha}\over \partial x^{\alpha}}{\partial \over \partial \epsilon}
\right) 
\inn \left( \sum_{\alpha=0}^{n-1} p^{\alpha}{\partial g\over \partial x^{\alpha}}\omega
- g{\partial \over \partial y}\inn \Omega\right) \\
& = & \displaystyle  g\sum_{\alpha=0}^{n-1}f^{\alpha}\omega_{\alpha},
\end{array}$$
and $\{Q^f,Q^{f'}\} = \{P_g,P_{g'}\} = 0$.
We now integrate the $\mathfrak{p}$-brackets on a constant time slice $S_t\subset \Gamma$. We 
immediately see that

$$\int_{S_t}\{P_g,Q^f\} =   \int_{S_t}g\; f^0\omega_0
= \{\pi_g(t),\Phi^f(t)\} = \left\{ \int_{S_t}P_g,\int_{S_t}Q^f\right\}$$
and we recover (\ref{trans}). Second,

$$\int_{S_t}\{\eta_0,Q^f\} = \int_{S_t}\sum_{\alpha=0}^{n-1}f^{\alpha}
{\partial {\cal H}\over \partial p^{\alpha}}\omega_0
- \sum_{\alpha=1}^{n-1}f^{\alpha}{\partial \phi\over \partial x^{\alpha}} \omega_0.$$
Third,

$$\int_{S_t}\{\eta_0,P_g\} =  \int_{S_t}-g{\partial {\cal H}\over \partial y}\omega_0
- \sum_{\alpha=1}^{n-1}g{\partial p^{\alpha}\over \partial x^{\alpha}}\omega_0.
$$
Now let us assume that $\Gamma$ is the graph of a solution of the Hamilton equations
(\ref{hamilton3}) or (\ref{weyl}).
Since then ${\partial \phi\over \partial x^{\alpha}}
= {\partial {\cal H}\over \partial p^{\alpha}}$ along $\Gamma$,

$$\int_{S_t}\{\eta_0,Q^f\} = \int_{S_t}f^0{\partial \phi\over \partial t}\omega_0,$$
and because of $-{\partial {\cal H}\over \partial y}
- \sum_{\alpha=1}^{n-1}{\partial p^{\alpha}\over \partial x^{\alpha}} =
{\partial p^0\over \partial t}$,

$$\int_{S_t}\{\eta_0,P_g\} = \int_{S_t}g{\partial p^0\over \partial t}\omega_0.$$
We conclude that

$${d\over dt}\int_{S_t}Q^f =
{d\over dt}\Phi^f(t) = \int_{S_t}f^0{\partial \phi\over \partial t}\omega_0
+{\partial f^0\over \partial t}\phi\omega_0
= \int_{S_t}\{\eta_0,Q^f\} + \Phi^{\partial f/\partial t}(t) $$
and

$${d\over dt}\int_{S_t}P_g =
{d\over dt}\Pi_g(t) = \int_{S_t}g{\partial p^0\over \partial t}\omega_0 +
{\partial g\over \partial t}p^0\omega_0
= \int_{S_t}\{\eta_0,P_g\} + \Pi_{\partial g/\partial t}(t).$$
This has to be compared with the usual canonical equations for fields:

$${d\over dt}\int_{S_t}Q^f =
\left\{\int_{S_t}\eta_0,\int_{S_t}Q^f\right\} + \Phi^{\partial f/\partial t}(t)
\quad \hbox{and}\quad
{d\over dt}\int_{S_t}P_g =
\left\{\int_{S_t}\eta_0,\int_{S_t}P_g\right\} + \Pi_{\partial g/\partial t}(t).$$

\subsection{An alternative dynamical formulation using $\mathfrak{p}$-brackets}

\noindent We can also define the $\mathfrak{p}$-bracket of a $n$-form with forms
$Q^f$ or $P_g$ as given by (\ref{3.3.f}) and (\ref{3.3.QP}).
If $\psi$ is such a $n$-form, 

$$\{ \psi, Q^f\} := -\xi_{Q^f}\inn d\psi\quad
\hbox{and}\quad \{ \psi, P_g\} := -\xi_{P_g}\inn d\psi,$$
where (\ref{3.3.A}) and (\ref{3.3.B}) have been used.
For instance, let us apply this definition to the $n$-form
$\eta:= {\cal H}(q,p)\omega - \theta$. We compute that

$$\{ \eta, Q^f\} = \sum_{\alpha=0}^{n-1}{\partial {\cal H}\over \partial p^{\alpha}}f^{\alpha}\omega
- \sum_{\alpha=0}^{n-1}f^{\alpha}dy\wedge \omega_{\alpha}$$
and

$$\{ \eta, P_g\} = - g{\partial {\cal H}\over \partial y}\omega - \sum_{\alpha=0}^{n-1}g\; dp^{\alpha}\wedge \omega_{\alpha}.$$
Now we integrate these $\mathfrak{p}$-brackets on
$\Gamma_{t_1}^{t_2}:= \{(q,p)\in \Gamma/ t_1<t<t_2\}$ and we
still assume that $\Gamma$ is the graph of a solution of
the Hamilton equations (\ref{weyl}): using these equations we find that

$$\int_{\Gamma_{t_1}^{t_2}}\{ \eta, Q^f\} = \int_{\Gamma_{t_1}^{t_2}}\{ \eta, P_g\} = 0.$$

On the other hand, we may compute also

$$\{ {\cal H}\omega, Q^f\} = \sum_{\alpha=0}^{n-1}f^{\alpha}{\partial {\cal H}\over \partial p^{\alpha}}\omega
+ \sum_{\alpha=0}^{n-1}y{\partial f^{\alpha}\over \partial x^{\alpha}}\omega,$$
and integrating by parts on $\Gamma_{t_1}^{t_2}$,

$$\begin{array}{ccl}
\displaystyle \int_{\Gamma_{t_1}^{t_2}}\{ {\cal H}\omega, Q^f\} & = &\displaystyle 
\int_{\partial \Gamma_{t_1}^{t_2}}\phi \sum_{\alpha=0}^{n-1}f^{\alpha}\omega_{\alpha}
+ \int_{\Gamma_{t_1}^{t_2}}\sum_{\alpha=0}^{n-1}f^{\alpha}\left( {\partial {\cal H}\over \partial p^{\alpha}}
- {\partial \phi\over \partial x^{\alpha}}\right) \omega\\
& = & \displaystyle \int_{\partial \Gamma_{t_1}^{t_2}}Q^f = \int_{S_{t_2}}Q^f - \int_{S_{t_1}}Q^f.
\end{array}$$
Similarly we find that

$$\{ {\cal H}\omega, P_g\} = \sum_{\alpha=0}^{n-1}p^{\alpha}{\partial g\over \partial x^{\alpha}}\omega
- g{\partial {\cal H}\over \partial y}\omega,$$
and thus

$$\begin{array}{ccl}
\displaystyle \int_{\Gamma_{t_1}^{t_2}}\{ {\cal H}\omega, P_g\} & = &\displaystyle 
\int_{\partial \Gamma_{t_1}^{t_2}}\sum_{\alpha=0}^{n-1}gp^{\alpha}\omega_{\alpha}
- \int_{\Gamma_{t_1}^{t_2}}g\left( {\partial {\cal H}\over \partial y}
+ \sum_{\alpha=0}^{n-1}{\partial p^{\alpha}\over \partial x^{\alpha}}\right) \omega \\
& =& \displaystyle \int_{\partial \Gamma_{t_1}^{t_2}}P_g = \int_{S_{t_2}}P_g - \int_{S_{t_1}}P_g.
\end{array}$$
We are tempted to conclude that

$${\bf d}Q^f = \{ {\cal H}\omega, Q^f\}\quad \hbox{and}\quad 
{\bf d}P_g = \{ {\cal H}\omega, P\},$$
where ${\bf d}$ is the differential along a graph $\Gamma$ of a solution of the Hamilton equations
(\ref{hamilton3}). This precisely will be proven in the next section.

\section{Poisson $\mathfrak{p}$-brackets on $(n-1)$-forms and more}
We have seen on some example that the Poisson bracket algebra of the classical field theory
can actually be derived from brackets on $(n-1)$-forms which are integrated
on constant time slices. Actually these constructions can be generalized in
several ways.

\subsection{Internal and external $\mathfrak{p}$-brackets}
We turn back to ${\cal M} = \Lambda^nT^{\star}({\cal X}\times {\cal Y})$ and
to the notation of the previous Section.
Let $\Gamma({\cal M},\Lambda^{n-1}T^{\star}{\cal M})$ be the set of smooth $(n-1)$-forms
on ${\cal M}$.  We consider the subset $\mathfrak{P}^{n-1}{\cal M}$ of
$\Gamma({\cal M},\Lambda^{n-1}T^{\star}{\cal M})$ of
forms $a$ such that there exists a vector field $\xi_{a}=\Xi(a)$
which satisfies the property

$$da = - \xi_{a}\inn \Omega.$$
Obviously $\Xi({a})$ depends only on $a$ modulo closed forms and the
map $a\longmapsto \Xi({a})$ from $\mathfrak{P}^{n-1}{\cal M}$
to the set of vector fields induces a map on the quotient
$\mathfrak{P}^{n-1}{\cal M}/C^{n-1}({\cal M})$, where $C^{n-1}({\cal M})$
is the set of closed $(n-1)$-forms.
A property of vector fields $\Xi(a)$ is that there are infinitesimal
symmetries of $\Omega$, for

$${\cal L}_{\Xi(a)}\Omega = d\left( \Xi(a)\inn \Omega\right)
+ \Xi(a)\inn d\Omega = -d\circ da = 0.$$
We shall denote $\mathfrak{pp}{\cal M}$ the set of pataplectic vector fields, i.e.
vector fields $X$ such that $X\inn \Omega$ is exact. Clearly
$\Xi:\mathfrak{P}^{n-1}{\cal M}/C^{n-1}({\cal M})\longrightarrow \mathfrak{pp}{\cal M}$
is a vector space isomorphism.

Then we define the {\em internal $\mathfrak{p}$-bracket} on $\mathfrak{P}^{n-1}{\cal M}$ by

$$\{a,b\} := \Xi({b})\inn \Xi(a)\inn \Omega.$$
\begin{lemm}
For any $a,b\in \mathfrak{P}^{n-1}{\cal M}$,
$$d\{a,b\} = - [\Xi({a}),\Xi({b})]\inn \Omega.$$
\end{lemm}
{\bf Proof} Let $\xi_{a}=\Xi(a)$ and $\xi_{b}=\Xi(b)$. Then denoting
${\cal L}_{\xi_{a}}$ the Lie derivative with respect to $\xi_{a}$,

$$\begin{array}{ccl}
[\xi_{a},\xi_{b}] \inn \Omega & = & {\cal L}_{\xi_{a}}(\xi_{b})\inn \Omega\\
 & = &{\cal L}_{\xi_{a}}\left( \xi_{b}\inn \Omega\right) -
\xi_{b}\inn {\cal L}_{\xi_{a}}(\Omega)\\
 & = & d(\xi_{a}\inn \xi_{b}\inn \Omega) + \xi_{a}\inn d(\xi_{b}\inn \Omega)
- \xi_{b}\inn (d(\xi_{a}\inn \Omega) + \xi_{a}\inn d\Omega).
\end{array}$$
But since $d\Omega = d(\xi_{a}\inn \Omega) = d(\xi_{b}\inn \Omega) = 0$, we find that
$[\xi_{a},\xi_{b}]\inn \Omega = d(\xi_{a}\inn \xi_{b}\inn \Omega)
= -d\{a,b\}$. \bbox
We deduce from this Lemma that 
$\Xi(\{a,b\}) = [\Xi({a}),\Xi({b})]$
and hence the map\\
$\Xi:\mathfrak{P}^{n-1}{\cal M}/C^{n-1}({\cal M}) \longrightarrow \mathfrak{pp}{\cal M}$
is actually a Lie algebra isomorphism. Notice that the Jacobi identity for the internal
$\mathfrak{p}$-bracket modulo exact terms is a consequence of this isomorphism.\\

We can extend this definition: for any
$0\leq p\leq n$ the {\em external $\mathfrak{p}$-bracket} of a $p$-form
$a\in \Gamma({\cal M},\Lambda^pT^{\star}{\cal M})$
with a form $b\in \mathfrak{P}^{n-1}{\cal M}$ is

$$\{a,b\} = - \{b,a\}:= - \Xi(b)\inn da.$$
Of course this definition coincides with the previous one when
$a\in \mathfrak{P}^{n-1}{\cal M}$.\\

\noindent {\bf Examples of external $\mathfrak{p}$-brackets}
For any $a\in \mathfrak{P}^{n-1}{\cal M}$,

$$\{\theta,a\} = -\Xi(a)\inn d\theta = - \Xi(a)\inn \Omega
= da.$$
We can add that it is worthwhile to write in the external $\mathfrak{p}$-brackets
of observable forms
like $q^{\mu}$, $q^{\mu}dq^{\nu}$, etc ...
$$\begin{array}{ccl}
\{P_{i,g},q^{\mu}\} & = & \Xi(P_{i,g}) \inn dq^{\mu} = g \delta^{\mu}_i,\\
\{Q^{i,f},q^{\mu}\} & = & \Xi(Q^{i,f}) \inn dq^{\mu} = 0\\
\{P_{i,g},q^{\mu}dq^{\nu}\} & = & \Xi(P_{i,g}) \inn dq^{\mu}\wedge dq^{\nu} =
g\left(  \delta^{\mu}_idq^{\nu} - \delta^{\nu}_idq^{\mu}\right) .
\end{array}$$

\begin{theo}
Let $\Gamma$ be the graph in ${\cal M}$ of a solution of the Hamilton equations
(\ref{hamilton3}) and write ${\cal U}:x\longmapsto {\cal U}(x) = (x,u(x),p(x))$ the natural
parametrization of $\Gamma$. Then for any form $a\in \mathfrak{P}^{n-1}{\cal M}$,

$${\bf d}a = \{{\cal H}\omega,a\},$$
where ${\bf d}$ is the differential along $\Gamma$
(meaning that $da_{|\Gamma} = \{{\cal H}\omega,a\}_{|\Gamma}$).
\end{theo}
{\bf Proof} We choose an arbitrary open subset $D\in \Gamma$ and 
denoting $\xi_{a} = \Xi(a)$, we compute

$$\begin{array}{ccl}
\displaystyle \int_D\{{\cal H}\omega,a\} & = & \displaystyle 
- \int_D\xi_{a}\inn (d{\cal H}\wedge \omega)\\
& = & \displaystyle - \int_{{\cal U}^{-1}(D)}d{\cal H}\wedge \omega
\left( \xi_{a},{\partial {\cal U}\over \partial x^1},\dots ,
{\partial {\cal U}\over \partial x^n}\right) 
\omega\\
& = & \displaystyle - \int_{{\cal U}^{-1}(D)}\left[ d{\cal H}(\xi_{a})\omega
\left( {\partial {\cal U}\over \partial x^1},\dots ,
{\partial {\cal U}\over \partial x^n}\right) \right.\\
&  & \displaystyle
- \left. \sum_{\alpha = 1}^nd{\cal H}\left( {\partial {\cal U}\over \partial x^{\alpha}}\right)
\omega\left( {\partial {\cal U}\over \partial x^1},\dots ,
{\partial {\cal U}\over \partial x^{\alpha-1}},\xi_{\alpha},{\partial {\cal U}\over \partial x^{\alpha+1}},
\dots ,
{\partial {\cal U}\over \partial x^n}\right) \right] \omega \\
& = & \displaystyle - \int_{{\cal U}^{-1}(D)}\left[ d{\cal H}(\xi_{a})- \sum_{\beta =1}^n\xi_{a}^{\beta}
d{\cal H}\left( {\partial {\cal U}\over \partial x^{\beta}}\right) \right] \omega.
\end{array}$$

We use  equation (\ref{hamilton6}) and obtain

$$\begin{array}{ccl}
\displaystyle \int_D\{{\cal H}\omega,a\} & = & \displaystyle 
- \int_{{\cal U}^{-1}(D)}(-1)^n{\partial {\cal U}\over \partial x^1\dots \partial x^n}\inn
\Omega(\xi_{a})\omega\\
& = & \displaystyle 
- \int_{{\cal U}^{-1}(D)}\Omega\left( \xi_{a},
{\partial {\cal U}\over \partial x^1},\dots ,{\partial {\cal U}\over \partial x^n}\right) 
\omega\\
& = & \displaystyle 
- \int_D\xi_{a}\inn \Omega = \int_Dda.
\end{array}$$
And the Theorem follows. \bbox
Another way to state this result is that

\begin{equation}\label{stokes}
\int_D\{{\cal H}\omega,a\} = \int_{\partial D}a
\end{equation}
along any solution of (\ref{hamilton3}).\\

\subsection{Expression of the standard observable quantities}

\noindent These quantities are integrals of $(n-1)$-forms on hypersurfaces
which are thought as ``constant time slices'', the transversal dimension being
then considered as a local time. The target coordinates observables
\footnote{\label{position}comparing with the one-dimensional Hamiltonian formalism we can
see these target coordinates as generalizations of the position observables.}
are weighted integrals
of the value of the field and are induced by the {\em ``position'' $\mathfrak{p}$-forms}

$$Q^{i,f} := y^i\; \sum_{\alpha}f^{\alpha}(x) \omega_{\alpha} = y^i\; f\inn \omega,$$
where $f=\sum_{\alpha}f^{\alpha}(x){\partial \over \partial x^{\alpha}}$ is a
tangent vector field on ${\cal X}$ and
$\omega_{\alpha} = {\partial \over \partial x^{\alpha}}\inn \omega$. The ``momentum''
and ``energy'' observables are obtained from the {\em momentum form}

$$P_{\mu,g}^{\star}:=
g(x) {\partial \over \partial q^{\mu}}\inn (\theta-{\cal H}(q,p)\omega),$$
where $g$ is a smooth function on ${\cal X}$. Alternatively we may sometimes prefer to use
the $\mathfrak{p}$-forms

$$P_{\mu,g}:= g(x) {\partial \over \partial q^{\mu}}\inn \theta.$$
For $1\leq \mu=\alpha\leq n$,
$P_{\mu,g}^{\star} =: H_{\alpha,g}$ generates the components of the Hamiltonian
tensor but $P_{\alpha,g}$ (which is different from $P_{\alpha,g}^{\star}$)
does not in general.
However the restrictions of $P_{\mu,g}^{\star}$
and $P_{\mu,g}$ on the hypersurface ${\cal H}=0$ coincide so that if we work
on this hypersurface both forms can be used.
For $n+1\leq \mu=n+i\leq n+k$, $P_{\mu,g}^{\star} =P_{\mu,g} =: P_{i,g}$ generates the
momentum components
\footnote{ The advantage
of $P_{\mu,g}$ with respect to $P_{\mu,g}^{\star}$ is that
$P_{\mu,g}$ belongs to $\mathfrak{P}^{n-1}{\cal M}$ for all values of $\mu$.}.\\

To check that, we consider a parametrization
${\cal U}:x\longmapsto (x,u(x),p(x))$ of some graph $\Gamma$ and look at
the pull-back of these forms by ${\cal U}$. We write
${\cal U}^{\star}P_{\mu,g}^{\star} = \sum_{\beta}s^{\beta}\omega_{\beta}$, which implies
$s^{\beta}\omega= dx^{\beta}\wedge {\cal U}^{\star}P_{\mu,g}^{\star}$ and we compute
$$\begin{array}{ccl}
s^{\beta} & = & \displaystyle g(x)\left\langle p,{\partial q\over \partial x^1}\wedge \dots
\wedge {\partial q\over \partial x^{\beta-1}}\wedge
{\partial \over \partial q^{\mu}}\wedge {\partial q\over \partial x^{\beta+1}}
\wedge \dots \wedge {\partial q\over \partial x^n}\right\rangle
_{|z={\partial U\over \partial x}}
- g(x)\delta^{\beta}_{\mu}{\cal H}\\
& = &\displaystyle g(x){\partial \langle p,z\rangle \over \partial z^{\mu}_{\beta}}
_{|z={\partial U\over \partial x}}
- g(x)\delta^{\beta}_{\mu}{\cal H}.
\end{array}$$

Hence we find that

$${\cal U}^{\star}H_{\alpha,g} = -g(x)\sum_{\beta}H^{\beta}_{\alpha}(q(x),p(x))
\omega_{\beta} = g(x)\sum_{\beta}S^{\beta}_{\alpha}(x,u(x),du(x))\omega_{\beta},$$
$${\cal U}^{\star}P_{i,g} = g(x)\sum_{\beta}
{\partial \langle p,z\rangle \over \partial z^i_{\beta}}_{|z={\partial U\over \partial x}}
\omega_{\beta} = g(x)\sum_{\beta}{\partial L\over \partial v^i_{\beta}}(x,u(x),du(x))
\omega_{\beta}.$$
We shall prove below that $P_{\mu,g}$ (and hence $P_{i,g}$) and $Q^{i,f}$ belong to
$\mathfrak{P}^{n-1}{\cal M}$.\\

\subsubsection{Larger classes of observable}

These forms, which are enough to translate most of the observable
studied in the usual field theory, are embedded in two more general classes of
observables the definition of which follows.\\

\noindent {\bf Generalised positions} (see the footnote \ref{position})
For each section of
$\Lambda^2T{\cal M}$ (i. e. a 2-vector field) of the form

$$\zeta:= \sum_{\mu_1<\dots <\mu_n}\sum_{\alpha}
\zeta^{\mu_{\alpha}}_{\mu_1\dots \mu_n}(q){\partial \over \partial p_{\mu_1\dots \mu_n}}
\wedge {\partial \over \partial q^{\mu_{\alpha}}},$$
we define the $(n-1)$-form

$$Q^{\zeta}:= \zeta\inn \Omega =
\sum_{\mu_1<\dots <\mu_n}\sum_{\alpha}
\zeta^{\mu_{\alpha}}_{\mu_1\dots \mu_n}{\partial \over \partial q^{\mu_{\alpha}}}\inn
{\partial \over \partial p_{\mu_1\dots \mu_n}}\inn \Omega. $$
An example is for
$\zeta = y^if(x){\partial \over \partial \epsilon}\wedge
{\partial \over \partial x^{\alpha}}$. It gives
$Q^{y^if(x){\partial \over \partial \epsilon}\wedge
{\partial \over \partial x^{\alpha}}} = Q^{i,f}$.
We denote $\mathfrak{P}_Q^{n-1}{\cal M}$ the set of such $(n-1)$-forms.\\

\noindent {\bf Generalised momenta} For each section of
$T({\cal X}\times{\cal Y})$, i. e. a vector field

$$\xi:= \sum_{\mu}\xi^{\mu}(q){\partial \over \partial q^{\mu}},$$
we define the $(n-1)$-form

$$P_{\xi}:= \xi\inn \theta =
\sum_{\mu}\xi^{\mu}{\partial \over \partial q^{\mu}}\inn \theta.$$
An example is for $\xi = g(x){\partial \over \partial q^{\mu}}$, then we
obtain $P_{g(x){\partial \over \partial q^{\mu}}} = P_{\mu,g}$.
We denote $\mathfrak{P}_P^{n-1}{\cal M}$ the set of such $(n-1)$-forms.

\begin{lemm}
All the $(n-1)$-forms defined above are in $\mathfrak{P}^{n-1}{\cal M}$, precisely
$$\begin{array}{ccl}
\Xi(Q^{\zeta}) & = & \displaystyle - \sum_{\mu_1<\dots <\mu_n}\sum_{\alpha}\sum_{\nu}
{\partial \zeta^{\nu}_{\mu_1\dots \mu_{\alpha-1}\nu\mu_{\alpha+1}\dots \mu_n}
\over \partial q^{\mu_{\alpha}}}
{\partial \over \partial p_{\mu_1\dots \mu_n}},\\
 & & \\
\Xi(P_{\xi}) & = & \displaystyle \xi - \sum_{\mu}\sum_{\nu}
{\partial \xi^{\mu}\over \partial q^{\nu}}\Pi^{\nu}_{\mu},
\end{array}$$
where

$$\Pi^{\nu}_{\mu}:= \sum_{\mu_1<\dots <\mu_n}\sum_{\alpha}
p_{\mu_1\dots \mu_{\alpha-1}\mu\mu_{\alpha+1}\dots \mu_n}
\delta^{\nu}_{\mu_{\alpha}}{\partial \over \partial p_{\mu_1\dots \mu_n}}$$
so that 

$$dq^{\nu}\wedge {\partial \over \partial q^{\mu}}\inn \theta = 
\Pi^{\nu}_{\mu}\inn \Omega.$$
\end{lemm}
{\bf Proof} Using the relation

$$dq^{\nu}\wedge\left( {\partial \over \partial q^{\mu_{\alpha}}}
\inn {\partial \over \partial p_{\mu_1\dots \mu_n}}\inn
\Omega\right)  =
{\partial \over \partial p_{\mu_1\dots \mu_{\alpha-1}\nu\mu_{\alpha+1}\dots \mu_n}}\inn
\Omega,$$
we obtain

$$\begin{array}{ccl}
dQ^{\zeta} & = & \displaystyle \sum_{\mu_1,\dots ,\mu_n}\sum_{\alpha}\sum_{\nu}
{1\over n!}
{\partial \zeta^{\mu_{\alpha}}_{\mu_1\dots  \mu_n}\over \partial q^{\nu}}
{\partial \over \partial p_{\mu_1\dots \mu_{\alpha-1}\nu \mu_{\alpha+1}\dots \mu_n}}
\inn \Omega\\
 & = & \displaystyle \sum_{\mu_1,\dots ,\mu_n}\sum_{\alpha}\sum_{\nu}
{1\over n!}
{\partial \zeta^{\nu}_{\mu_1\dots \mu_{\alpha-1}\nu \mu_{\alpha+1}\dots \mu_n}
\over \partial q^{\mu_{\alpha}}}
{\partial \over \partial p_{\mu_1\dots \mu_n}}
\inn \Omega.
\end{array}$$
And the expression for $\Xi(Q^{\zeta})$ follows.\\

\noindent Next we write

$$dP_{\xi} = \sum_{\mu}\sum_{\nu}{\partial \xi^{\mu}\over \partial q^{\nu}}
dq^{\nu}\wedge {\partial \over \partial q^{\mu}}\inn \theta 
- \sum_{\mu}\xi^{\mu}{\partial \over \partial q^{\mu}}\inn \Omega$$
and we conclude by computing
$dq^{\nu}\wedge {\partial \over \partial q^{\mu}}\inn \theta$, indeed

$$\begin{array}{ccl}
\displaystyle dq^{\nu}\wedge {\partial \over \partial q^{\mu}}\inn \theta
& = & \displaystyle \sum_{\mu_1<\dots <\mu_n}\sum_{\alpha}
p_{\mu_1\dots \mu_n}\delta^{\mu_{\alpha}}_{\mu}
dq^{\mu_1}\wedge \dots \wedge dq^{\mu_{\alpha-1}}\wedge dq^{\nu}\wedge
dq^{\mu_{\alpha+1}}\wedge \dots \wedge dq^{\mu_n}\\
 & = &  \displaystyle \sum_{\mu_1<\dots <\mu_n}\sum_{\alpha}
p_{\mu_1\dots \mu_{\alpha-1}\mu\mu_{\alpha+1}\dots \mu_n}\delta^{\nu}_{\mu_{\alpha}}
dq^{\mu_1}\wedge \dots \wedge dq^{\mu_n}\\
 & = &  \displaystyle \sum_{\mu_1<\dots <\mu_n}\sum_{\alpha}
p_{\mu_1\dots \mu_{\alpha-1}\mu\mu_{\alpha+1}\dots \mu_n}\delta^{\nu}_{\mu_{\alpha}}
{\partial \over \partial p_{\mu_1\dots \mu_n}}\inn \Omega.
\end{array}$$
Hence we deduce the result on $P_{\xi}$.\bbox 

\noindent {\bf Poisson $\mathfrak{p}$-brackets}\\

\noindent We are now in position to compute the $\mathfrak{p}$-brackets of these
forms. The results are summarized in the following Proposition.

\begin{prop}
The $\mathfrak{p}$-brackets of forms in $\mathfrak{P}_Q^{n-1}{\cal M}$ and
$\mathfrak{P}_P^{n-1}{\cal M}$
are the following

$$\begin{array}{ccl}
\{Q^{\zeta}, Q^{\tilde{\zeta}}\} & = & 0\\
 & & \\
\{P_{\xi}, P_{\tilde{\xi}}\} & = & 
P_{[\xi,\tilde{\xi}]}+ d(\tilde{\xi}\inn \xi \inn \theta)\\
 & & \\
\{P_{\xi}, Q^{\zeta}\} & = &\displaystyle 
\sum_{\mu_1<\dots <\mu_n}\sum_{\alpha}\sum_{\mu}\sum_{\nu}\xi^{\mu}
{\partial \zeta^{\nu}_{\mu_1\dots \mu_{\alpha-1}\nu \mu_{\alpha-1}\dots \mu_n}
\over \partial q^{\mu_{\alpha}}}
{\partial \over \partial q^{\mu}}\inn
dq^{\mu_1}\wedge \dots \wedge dq^{\mu_n}.
\end{array}$$
\end{prop}
{\bf Proof} These results are all straighforward excepted for $\{P_{\xi}, P_{\tilde{\xi}}\}$,

\begin{equation}\label{4.2.1.lie}
\begin{array}{ccl}
{\cal L}_{\Xi(P_{\xi})}(\theta) & = & \Xi(P_{\xi})\inn d\theta + d(\Xi(P_{\xi})\inn \theta)\\
 & = & \Xi(P_{\xi})\inn \Omega + dP_{\xi} = 0,
\end{array}
\end{equation}
so that $\Xi(P_{\xi})$ may be viewed as the extension of $\xi$ to a vector
field leaving $\theta$ invariant. Now we deduce that

$$\begin{array}{ccl}
[\xi, \tilde{\xi}]\inn \theta & = & [\Xi(P_{\xi}), \Xi(P_{\tilde{\xi}})]\inn \theta\\
 & = & {\cal L}_{\Xi(P_{\xi})}(\Xi(P_{\tilde{\xi}}))\inn \theta\\
 & = & {\cal L}_{\Xi(P_{\xi})}(\Xi(P_{\tilde{\xi}})\inn \theta)
- \Xi(P_{\tilde{\xi}})\inn {\cal L}_{\Xi(P_{\xi})}\theta\\
 & = & \Xi(P_{\xi})\inn d(\Xi(P_{\tilde{\xi}})\inn \theta) +
d (\Xi(P_{\xi})\inn \Xi(P_{\tilde{\xi}})\inn \theta)
- \Xi(P_{\tilde{\xi}})\inn 0\\
 & = & \Xi(P_{\xi})\inn dP_{\tilde{\xi}} + d(\xi\inn \tilde{\xi}\inn \theta)\\
 & = & - \Xi(P_{\xi})\inn \Xi(P_{\tilde{\xi}})\inn \Omega + d(\xi\inn \tilde{\xi}\inn \theta)\\
 & = & \{P_{\xi},P_{\tilde{\xi}}\} - d(\tilde{\xi}\inn \xi\inn \theta)
\end{array}$$
And the result follows. \bbox

\subsubsection{Back to the standard observables}

As an application of the previous results we can express the pataplectic vector
fields associated to $Q^{i,f}$ and $P_{\mu,g}$ and their $\mathfrak{p}$-brackets.
For that purpose, it is useful to introduce other notations:

$$\begin{array}{ccl}
\epsilon & := & p_{1\dots n}\\
p^{\alpha}_i & := & p_{1\dots (\alpha-1)(n+i)(\alpha+1)\dots n}\\
p^{\alpha_1\alpha_2}_{i_1i_2} & := &
p_{1\dots (\alpha_1-1)(n+i_1)(\alpha_1+1)\dots 
(\alpha_2-1)(n+i_2)(\alpha_2+1)\dots n}\\
& & \hbox{etc}\dots 
\end{array}$$
and
$$\begin{array}{ccl}
\omega^i_{\alpha} & := & dy^i\wedge \left( {\partial \over \partial x^{\alpha}}
\inn \omega\right) =: (dy^i\wedge \partial _{\alpha})\inn \omega\\
\omega^{i_1i_2}_{\alpha_1\alpha_2} & := & (dy^{i_1}\wedge \partial _{\alpha_1})\inn 
(dy^{i_2}\wedge \partial _{\alpha_2})\inn  \omega\\
& & \hbox{etc}\dots 
\end{array}$$
in such a way that

$$\theta = \epsilon\;\omega  + \sum_{p=1}^n{1\over p!^2}
\sum_{i_1,\dots ,i_p;\alpha_1,\dots ,\alpha_p}
p^{\alpha_1\dots \alpha_p}_{i_1\dots i_p}
\omega^{i_1\dots i_p}_{\alpha_1\dots \alpha_p}.$$
(Notice that the Weyl theory corresponds to the assumption that
$p^{\alpha_1\dots \alpha_p}_{i_1\dots \i_p}=0$, $\forall p\geq 2$.)
We have

$$\begin{array}{ccl}
dQ^{i,f} & = & \displaystyle \sum_{\alpha}f^{\alpha}{\partial \over \partial p_i^{\alpha}}\inn \Omega
+ y^i\sum_{\alpha}{\partial f^{\alpha}\over \partial x^{\alpha}}
{\partial \over \partial \epsilon}\inn \Omega,\\
 & & \\
dP_{\mu,g} & = & \displaystyle \sum_{\alpha}{\partial g\over \partial x^{\alpha}}
\Pi^{\alpha}_{\mu}\inn \Omega -
g {\partial \over \partial q^{\mu}}\inn \Omega,
\end{array}$$
where

$$\begin{array}{ccl}
\Pi^{\alpha}_{\beta} & = & \displaystyle 
\delta^{\alpha}_{\beta}\epsilon{\partial \over \partial \epsilon}
+ \sum_{p=1}^{n}{1\over p!^2}\sum_{i_1,\dots ,i_p;\alpha_1,\dots ,\alpha_p}
\left( p^{\alpha_1\dots \alpha_p}_{i_1\dots i_p}\delta^{\alpha}_{\beta}
- \sum_{j=1}^p
p{^{\alpha_1\dots \alpha_{j-1}}_{i_1\dots i_{j-1}}}
{^{\alpha}_{i_j}}
{^{\alpha_{j+1}\dots \alpha_n}_{i_{j+1}\dots i_n}}
\delta^{\alpha_j}_{\beta}\right)

{\partial \over \partial p^{\alpha_1\dots \alpha_p}_{i_1\dots i_p}}\\
\Pi^{\alpha}_{n+i} & = &\displaystyle 
\sum_{p=0}^{n-1}{1\over p!^2}\sum_{i_1,\dots ,i_p;\alpha_1,\dots ,\alpha_p}
p^{\alpha\alpha_1\dots \alpha_p}_{ii_1\dots i_p}
{\partial \over \partial p^{\alpha_1\dots \alpha_p}_{i_1\dots i_p}}.
\end{array}$$
The pataplectic vector fields are

$$\begin{array}{ccl}
\Xi(Q^{i,f}) & = & \displaystyle - \sum_{\alpha}
f^{\alpha}{\partial \over \partial p_i^{\alpha}}
- y^i\sum_{\alpha}{\partial f^{\alpha}\over \partial x^{\alpha}}
{\partial \over \partial \epsilon}\\
\Xi(P_{\mu,g})&  = & \displaystyle g {\partial \over \partial q^{\mu}} -
\sum_{\alpha}{\partial g\over \partial x^{\alpha}} \Pi^{\alpha}_{\mu}.
\end{array}$$
Finally by using Proposition 1, the Poisson $\mathfrak{p}$-brackets will be

$$\begin{array}{ccl}
\{ Q^{i,f},Q^{j,\tilde{f}}\} & = & 0,\\
 & & \\
\{ P_{i,g},P_{j,\tilde{g}}\} & = & \displaystyle
d\left( g\tilde{g}{\partial \over \partial y^j}\inn 
{\partial \over \partial y^i}\inn \theta\right) ,\\
 & & \\
\{ P_{i,g},Q^{j,f}\} & = & \displaystyle
\delta^j_i
\sum_{\alpha}f^{\alpha}g\omega_{\alpha}.
\end{array}$$

\noindent Hence if $g$ and $\tilde{g}$ have compact support, we
obtain that on any submanifold $S$ of dimension $n-1$ without boundary,

$$\int_S\{ Q^{i,f},Q^{j,\tilde{f}}\} = \int_S \{ P_{i,g},P_{j,\tilde{g}}\} = 0
\quad \hbox{and}\quad
\int_S\{ P_{i,g},Q^{j,f}\} = \delta^j_i\int_S\sum_{\alpha}f^{\alpha}g\omega_{\alpha}$$

\subsection{The $\omega$-bracket}
The $\mathfrak{p}$-brackets defined above does not allow us to express the dynamics
of an observable which is not in $\mathfrak{P}^{n-1}{\cal M}$. Here is a
construction {\em ad hoc} of a bracket which induces the dynamics of
forms of arbitrary degree. In contrast with the $\mathfrak{p}$-bracket, which depends only on
$\Omega$ the following bracket relies also on the volume form $\omega$.

Let $X$ be some section of the bundle $\Lambda^nT{\cal M}\longrightarrow {\cal X}$,
such that $X$ is Hamiltonian everywhere, i. e. $(-1)^nX\inn \Omega = d{\cal H}$
mod ${\cal I}$.
For any integer $1\leq p\leq n$ and any form $\lambda\in \Lambda^{p-1}T^{\star}{\cal M}$
we define $X\sharp \lambda\in \Lambda^pT^{\star}{\cal M}$ by

\begin{equation}\label{xdieze}
X\sharp \lambda:= \sum_{\alpha_1<\dots <\alpha_{n-p}}
X\inn (dx^{\alpha_1\dots \alpha_{n-p}}\wedge d\lambda)\;
\partial_{\alpha_1\dots \alpha_{n-p}}\inn \omega,
\end{equation}
where $\partial_{\alpha_1\dots \alpha_{n-p}}:=
{\partial \over \partial x^{\alpha_1}}\wedge \dots \wedge
{\partial \over \partial x^{\alpha_{n-p}}}$,
$\partial_{\alpha_1\dots \alpha_{n-p}}\inn \omega(V_1,\dots ,V_p) = 
\omega\left( {\partial \over \partial x^{\alpha_1}}, \dots ,
{\partial \over \partial x^{\alpha_{n-p}}},V_1,\dots ,V_p\right)$ and
$dx^{\alpha_1\dots \alpha_{n-p}}:= dx^{\alpha_1}\wedge \dots \wedge dx^{\alpha_{n-p}}$
\footnote{\label{omegacro}we could meet situations (as in Section 5) where the volume form on ${\cal X}$
has been replaced by $\tilde{\omega}:= gdx^1\wedge \dots \wedge dx^n = g\omega$, where $g$ is a smooth positive function
on ${\cal X}$. This lead to replacing the Cartan form by $\tilde{\theta}:= g\theta$ and the pataplectic form
by $\tilde{\Omega}:= d(g\Omega)$. Then we can define $\tilde{X}\tilde{\sharp}\lambda$ by the same formula
as (\ref{xdieze}), where $\omega$ and $X$ are replaced by $\tilde{\omega}$ and $\tilde{X}$ respectively.
And we can check that $\tilde{X}\tilde{\sharp}\lambda = X\sharp \lambda$.}.

\begin{defi}
The $(p-1)$-form $\lambda$ is {\em admissible} if and only if, for any ${\cal H}$
and for any ${\cal H}$-Hamiltonian $n$-vector field $X$, $X\sharp \lambda$ does not
depend on the choice of $X$, but only on ${\cal H}$. If $\lambda$ is
admissible, we denote

$$\{{\cal H}\omega,\lambda\}_{\omega}:= X\sharp \lambda,$$
and name this a $\omega$-bracket.
\end{defi}

\noindent {\bf Examples} The 0-form $y^i$ and the 1-form $y^idy^j$ are
admissible and

$$\{{\cal H}\omega,y^i\}_{\omega} = \sum_{\alpha}{\partial {\cal H}\over \partial p^{\alpha}_i}
dx^{\alpha},$$
$$\{{\cal H}\omega,y^idy^j\}_{\omega} = \sum_{\alpha<\beta}
{\partial {\cal H}\over \partial p^{\alpha\beta}_{ij}}
dx^{\alpha}\wedge dx^{\beta}.$$
More generally, all forms in $\Lambda^{\star}T^{\star}({\cal X}\times {\cal Y})$
(``space-time'' and ``position'' observables) are admissible. We shall also see
below that the forms $Q^{i,f}$ and $P_{i,g}$ in $\mathfrak{P}^{n-1}{\cal M}$ are admissible
but not $P_{\alpha,g}$.
\begin{lemm}
Let $\Gamma$ be the graph of $x\longmapsto (q(x),p(x))$, a solution of the
Hamilton equations (\ref{hamilton7}) and let $\lambda$ be some
admissible $p$-form. Then $d\lambda$ coincides with
$\{{\cal H}\omega,\lambda\}_{\omega}$ along $\Gamma$, i. e.

$$d\lambda_{|\Gamma} = \{{\cal H}\omega,\lambda\}_{\omega|\Gamma}.$$
\end{lemm}
{\bf Proof} Let us denote $X = {\partial (q,p)\over \partial (x^1,\dots ,x^n)}$.
For $p = n-1$ this identity is obvious because

\begin{equation}\label{plem}
d\lambda_{|\Gamma} = (X\inn d\lambda )\omega = X\sharp \lambda.
\end{equation}
For $p<n-1$ we have, using (\ref{plem}),

$$\begin{array}{ccl}
d\lambda_{|\Gamma} & = & \displaystyle
\sum_{\alpha_1<\dots <\alpha_{n-p}} \partial_{\alpha_1\dots \alpha_{n-p}}\inn
(dx^{\alpha_1\dots \alpha_{n-p}}\wedge d\lambda_{|\Gamma})\\
 & = & \displaystyle
(-1)^{n-p}\sum_{\alpha_1<\dots <\alpha_{n-p}} \partial_{\alpha_1\dots \alpha_{n-p}}\inn
d(dx^{\alpha_1\dots \alpha_{n-p}}\wedge \lambda)_{|\Gamma}\\
 & =& \displaystyle
(-1)^{n-p}\sum_{\alpha_1<\dots <\alpha_{n-p}} \partial_{\alpha_1\dots \alpha_{n-p}}\inn
\left( X\inn d(dx^{\alpha_1\dots \alpha_{n-p}}\wedge \lambda)\omega\right) \\
 & =& \displaystyle
\sum_{\alpha_1<\dots <\alpha_{n-p}}
\left( \partial_{\alpha_1\dots \alpha_{n-p}}\inn \omega\right) 
(X\inn (dx^{\alpha_1\dots \alpha_{n-p}}\wedge d\lambda))\\
 & = & X\sharp \lambda_{|\Gamma}.
\end{array}$$
This achieves the proof. \bbox

Natural problems are to characterize the $(n-1)$-forms
in $\mathfrak{P}^{n-1}{\cal M}$ which
are admissible and
to compare the $\mathfrak{p}$-bracket and the $\omega$-bracket in cases where
they exist simultaneously. The answers are in the following.

\begin{prop}
{\em (i)} A form $a$ in $\mathfrak{P}^{n-1}{\cal M}$ is admissible if and
only if
\begin{equation}\label{ad+p}
dx^{\beta}(\Xi(a)) = 0,\quad \forall \beta = 1,\dots ,n.
\end{equation}
(As a consequence, examples of such forms are $Q^{i,f}$, $P_{i,g}$ but
not $P_{\alpha,g}$ \footnote{Forms like
$g(x){\partial \over \partial x^{\alpha}}\inn \theta$
are not admissible. However for any $a\in \mathfrak{P}^{n-1}$ and for any
solution $x\longmapsto (q(x),p(x))$
of the Hamilton equation the restriction of $\{{\cal H}\omega,a\}$ on
the graph coincide with the restriction of
${\partial (q,p)\over \partial (x^1,\dots ,x^n)}\sharp a$.}.)\\
{\em (ii)} For any admissible form $a\in \mathfrak{P}^{n-1}{\cal M}$,

\begin{equation}\label{p=ad}
\{{\cal H}\omega,a\} =\{{\cal H}\omega,a\}_{\omega}.
\end{equation}
\end{prop}
{\bf Proof} (i) Let $a\in \mathfrak{P}^{n-1}{\cal M}$, assume that $a$
is admissible and denote $\xi_{a}=\Xi(a)$.
Choose any decomposable ${\cal H}$-Hamiltonian
$X=X_1\wedge \dots \wedge X_n$, then

$$\begin{array}{ccl}
X\sharp a & = & (X\inn da)\omega = -(X\inn \xi_{a}\inn \Omega)\omega\\
 & = & -(-1)^n(\xi_{a}\inn X\inn \Omega)\omega\\
 & = & -\xi_{a} \inn \left( d{\cal H} - \sum_{\beta}
d{\cal H}\left( X_{\beta}\right)dx^{\beta}\right)
\omega ,
\end{array}$$
where we have used Lemma 1 for the last equality. We set $\xi_{a} = \sum_{\beta}\xi_{a}^{\beta}\partial_{\beta} + \sum_i\xi_{a}^i\partial_i
+ \sum_{\mu_1<\dots <\mu_n}\xi_{a,\mu_1\dots \mu_n}\partial^{\mu_1\dots \mu_n}$,
then we see that

$$X\sharp a =  \sum_{\beta}\xi_{a}^{\beta}
\left( {\partial {\cal H}\over \partial x^{\beta}} +
\sum_i{\partial {\cal H}\over \partial y^i}X^i_{\beta}
+ \sum_{\mu_1<\dots <\mu_n}
{\partial {\cal H}\over \partial p_{\mu_1\dots \mu_n}}
X_{\beta,\mu_1\dots \mu_n}\right)
\omega -(\xi_{a}\inn d{\cal H})\omega .$$

Since $X_{\beta,\mu_1\dots \mu_n}$ depends on the choice of $X$, we conclude that
we must have $\xi_{a}^{\beta} = 0$, i. e. (\ref{ad+p}) holds. Conversely if
(\ref{ad+p}) is true, then for any ${\cal H}$-Hamiltonian $n$-vector $X$
(not necessarily  decomposable)
$X\sharp a =-(-1)^n(\xi_{a}\inn X\inn \Omega)\omega$,
and according to (\ref{hamilvec1}) we deduce that
$X\sharp a =-\left( \xi_{a}\inn d{\cal H}\right) \omega$, an expression which
does not depend on the choice of $X$.\\

\noindent (ii) A consequence of the above calculation is that if
$a\in \mathfrak{P}^{n-1}{\cal M}$
is admissible then

$$\{{\cal H}\omega,a\}_{\omega} = -(\xi_{a}\inn d{\cal H})\omega.$$
Now (\ref{p=ad}) follows easily from
 
$$\{{\cal H}\omega,a\} = -\xi_{a}\inn d({\cal H}\omega) =
-\xi_{a}\inn d{\cal H}\wedge \omega$$
and condition (\ref{ad+p}). \bbox

\begin{rema}
It appears that it will be interesting to study solutions of the
Hamilton equations with the constraint ${\cal H} = 0$. This
is possible, because of the freedom left in the Legendre correspondance, thanks to
the parameter $\epsilon$.
The advantage is that then the energy-momentum observables are
described by $P_{a,g}$ which 
belongs to $\mathfrak{P}^{n-1}{\cal M}$.
\end{rema}

\subsection{Noether theorem}

It is natural to relate the Noether theorem
to the pataplectic structure.

Let $\xi$ be a tangent vector field on ${\cal X}\times {\cal Y}$,
$\xi$ will be an infinitesimal symmetry of the variational problem if

$${\cal L}_{\Xi(P_{\xi})}\left( \theta - {\cal H}\omega\right) = 0,$$
since then the integral $\int_{\Gamma}\theta - {\cal H}\omega$ is invariant
under the action of the flow of $\xi$.
Then for any solution $x\longmapsto (U(x),p(x))$, of the Hamilton equations,
the form $P^{\star}_{\xi}$ is closed along the
graph of this solution. This means that if $\Gamma$ is the graph of $(U,p)$, 

$$dP^{\star}_{\xi |\Gamma} = d\left( \xi \inn (\theta - {\cal H}\omega) \right) _{|\Gamma}
= 0.$$
This is a direct consequence of Theorem 2 and of the following calculation.

\begin{lemm}
For any section $\xi$ of $\Gamma({\cal X}\times {\cal Y}, T({\cal X}\times {\cal Y}))$,
we have the relation

\begin{equation}\label{noether}
\{{\cal H}\omega, P_{\xi}\} = 
{\cal L}_{\Xi(P_{\xi})}\left( \theta - {\cal H}\omega\right)
+ d\left( \xi \inn {\cal H}\omega \right) .
\end{equation}
\end{lemm}

{\bf Proof} Using the definition of $\{{\cal H}\omega, P_{\xi}\}$, we have

$$\begin{array}{ccl}
{\cal L}_{\Xi(P_{\xi})}\left( \theta - {\cal H}\omega\right) & = &
\Xi(P_{\xi})\inn (d\theta - d{\cal H}\wedge \omega )
+d\left( \Xi(P_{\xi})\inn (\theta -{\cal H}\omega)\right) \\
 & = & \Xi(P_{\xi})\inn \Omega - \Xi(P_{\xi})\inn d{\cal H}\wedge \omega
 + d\left( \xi\inn \theta -\xi\inn {\cal H}\omega\right) \\
  & = & -dP_{\xi} + \{{\cal H}\omega,P_{\xi}\} +
  d\left( P_{\xi} -\xi\inn {\cal H}\omega\right) ,
\end{array}$$
and the result follows. \bbox
\begin{rema}
As a consequence of these observations it is clear that on the submanifold ${\cal H} = 0$, the set
of Noether currents can be identified with $\mathfrak{P}_P^{n-1}{\cal M}$. So we can interpret the results of Proposition
1 concerning $\mathfrak{P}_P^{n-1}{\cal M}$ by saying that the set of Noether currents equipped with
the $\mathfrak{p}$-bracket is a representation modulo exact terms of the Lie algebra of vector
fields on ${\cal X}\times {\cal Y}$ with the Lie bracket. We recover thus various constructions
of brackets on Noether currents (see for instance \cite{Deligne-Freed}).
\end{rema}

\section{Examples}

We present here some examples from the mathematical Physics in order to illustrate
our formalism. We shall see that, by allowing variants of the above theory,
one can find formalisms which are more adapted to some special situations.

\subsection{Interacting scalar fields}
As the simplest example, consider a system of interacting scalar fields
$\{\phi^1,\dots ,\phi^k\}$ on an oriented (pseudo-)Riemannian manifold
$({\cal X},g)$. One should keep in mind that ${\cal X}$ is a four-dimensional
space-time and $g_{\alpha\beta}$ is a Minkowski metric.
These fields can be seen as a map $\phi$ from ${\cal X}$ to
$\Bbb{R}^k$ with its standard Euclidian structure. The metric $g$ on ${\cal X}$
induces a volume form which reads in local coordinates

$$\omega:= g\;dx^1\wedge \dots \wedge dx^n,\quad \hbox{ where }
g:= \sqrt{|\hbox{det}g_{\alpha\beta}(x)|}.$$
Let $V:{\cal X}\longrightarrow \Bbb{R}^k$ be the
interaction potential of the fields, then the Lagrangian density is

$$L(x,\phi,d\phi) := {1\over 2}g^{\alpha\beta}(x)
{\partial \phi^i\over \partial x^{\alpha}}
{\partial \phi_i\over \partial x^{\beta}} - V(\phi(x)).$$
Here $\phi_i=\phi^i$ and we assume that we sum over all repeated indices.
Alternatively one could work with the volume form being $dx^1\wedge \dots \wedge dx^n$
and the Lagrangian density being $gL$, in order to apply directly the
theory constructed in the previous sections. But we shall not choose this approach here
and use a variant which makes clear the covariance of the problem.

We restrict to the Weyl theory, i. e. we work on the the submanifold
${\cal M}_{\tiny{\hbox{Weyl}}}$, as in subsection 2.7. So we introduce the
momentum variables $\epsilon$ and $p^{\alpha}_i$ and we start from the
Cartan form

$$\theta = \epsilon\; \omega + p^{\alpha}_id\phi^i\wedge \omega_{\alpha},$$
where $\omega_{\alpha}:= \partial _{\alpha}\inn \omega$. But here $\omega_{\alpha}$
is not closed in general (because $g$ is not constant), so

$$\Omega = d\theta = d\epsilon\wedge \omega + dp^{\alpha}_i\wedge d\phi^i\wedge \omega_{\alpha}
- p^{\alpha}_i{1\over g}{\partial g\over \partial x^{\alpha}}d\phi^i\wedge \omega.$$
The Legendre transform is given by

$$p^{\alpha}_i = {\partial L\over \partial (\partial_{\alpha}\phi^i)} =
g^{\alpha\beta}{\partial \phi^i\over \partial x^{\beta}}
\Longleftrightarrow
{\partial \phi^i\over \partial x^{\alpha}} =
g_{\alpha\beta}p^{\beta}_i,$$
and the Hamiltonian is

$${\cal H}(x,\phi,p) = \epsilon + {1\over 2}g_{\alpha\beta}
p_i^{\alpha}p_i^{\beta} + V(\phi).$$

We use as conjugate variables the 0-forms $\phi^i$ and the $(n-1)$-forms

$$P_{i,f}:= f(x)p_i^{\alpha}\omega_{\alpha}
= f(x){\partial \over \partial \phi^i}\inn \theta\; \in
\mathfrak{P}^{n-1}{\cal M}_{\tiny{\hbox{Weyl}}}.$$
Taking account of the fact that $\omega_{\alpha}$ is not closed, one find

$$\Xi(P_{i,f}) = f{\partial \over \partial \phi^i}
- {\partial f\over \partial x^{\alpha}}p^{\alpha}_i{\partial \over \partial \epsilon}$$
and

$$\{P_{i,f},\phi^j\} = \Xi(P_{i,f})\inn d\phi^j = f\delta^j_i.$$
Also (see the footnote \ref{omegacro})

$$\begin{array}{ccl}
\{{\cal H}\omega,\phi^i\}_{\omega} & = & \displaystyle
{\partial {\cal H}\over \partial p^{\alpha}_i}
dx^{\alpha} = g_{\alpha\beta}p^{\beta}_idx^{\alpha}\\
&&\\
\{{\cal H}\omega,P_{i,f}\} & = & \displaystyle
- \Xi(P_{i,f})\inn d({\cal H}\omega)= 
\left( -f {\partial V\over \partial \phi^i}
+ {\partial f\over \partial x^{\alpha}}p^{\alpha}_i\right) \omega.
\end{array}$$
And the dynamical equations are that along the graph of a solution,

$$\begin{array}{ccccl}
{\bf d}\phi^i & = & \displaystyle
\{{\cal H}\omega,\phi^i\}_{\omega} & = &\displaystyle
g_{\alpha\beta}p^{\beta}_idx^{\alpha}\\
{\bf d}(fp^{\alpha}_i\omega_{\alpha}) & = & \displaystyle
\{{\cal H}\omega,P_{i,f}\} & = & \displaystyle
\left( -f {\partial V\over \partial \phi^i}
+ {\partial f\over \partial x^{\alpha}}p^{\alpha}_i\right) \omega.
\end{array}$$
The second equation gives

\begin{equation}\label{exa1}
{f\over g}\left( {\partial g\over \partial x^{\alpha}}p^{\alpha}_i
+g {\partial p^{\alpha}_i\over \partial x^{\alpha}}
+ g{\partial V\over \partial \phi^i}\right) = 0,
\end{equation}
while the first relation gives ${\partial \phi^i\over \partial x^{\alpha}}
= g_{\alpha\beta}p^{\beta}_i$. By substitution in (\ref{exa1}) we
find

$${1\over g}{\partial \over \partial x^{\alpha}}\left( 
g\;g^{\alpha\beta}{\partial \phi^i\over \partial x^{\beta}}\right) +
{\partial V\over \partial \phi^i} = 0,$$
the Euler-Lagrange equations of the problem.

\subsection{The conformal string theory}
We consider maps $u$ from a two-dimensional (pseudo-)Riemannian manifold
$({\cal X},g)$ with values in another (pseudo-)Riemannian manifold
$({\cal Y},h)$ of arbitrary dimension. The most general bosonic action for such maps
is ${\cal L}[u]:= \int_{\cal X}L(x,u,du)\omega$ with
$\omega:= g(x) dx^1\wedge dx^2$ and $g(x):= \sqrt{|\hbox{det}g_{\alpha\beta}(x)|}$
as before, and

$$L(x,u,du):= {1\over 2}\left( h_{ij}(u(x))g^{\alpha\beta}(x)
+ b_{ij}(u(x)){\epsilon^{\alpha\beta}\over g(x)}\right)
{\partial u^i\over \partial x^{\alpha}} 
{\partial u^j\over \partial x^{\beta}},$$
where $b:= \sum_{i<j}b_{ij}(y)dy^i\wedge dy^j$ is a given two-form on
${\cal Y}$ and $\epsilon^{12} = -\epsilon^{21} = 1$,
$\epsilon^{11} = \epsilon^{22} =0$. Hence

$${\cal L}[u] = \int_{\cal X}{1\over 2}h_{ij}(u)g^{\alpha\beta}(x)
{\partial u^i\over \partial x^{\alpha}} 
{\partial u^j\over \partial x^{\beta}}\omega +
u^{\star}b.$$
Setting

$$G^{\alpha\beta}_{ij}(x,y):= h_{ij}(y)g^{\alpha\beta}(x)
+ b_{ij}(y){\epsilon^{\alpha\beta}\over g(x)} = G^{\beta\alpha}_{ji}(x,y),$$
we see that $L(x,u,du) = {1\over 2}G^{\alpha\beta}_{ij}(x,u)
{\partial u^i\over \partial x^{\alpha}} 
{\partial u^j\over \partial x^{\beta}}$ and
the Euler-Lagrange equation for this functional is

\begin{equation}\label{string1}
{1\over g}{\partial \over \partial x^{\alpha}}\left( g\;G^{\alpha\beta}_{ij}(x,u(x))
{\partial u^j\over \partial x^{\beta}}\right)
= {\partial G^{\beta\gamma}_{jk}\over \partial y^i}
{\partial u^j\over \partial x^{\beta}}
{\partial u^k\over \partial x^{\gamma}}.
\end{equation}
More covariant formulations exists for the case $b=0$, which correspond
to the harmonic map equation or when the metric on ${\cal X}$ is
Riemannian using conformal coordinates and complex variables
(see \cite{Helein}). The Cartan-Poincar\'e form on ${\cal M}$
is

$$\theta:= \epsilon\; \omega + \sum_{\alpha,i}p^{\alpha}_i\omega^i_{\alpha}
+ \sum_{i<j}p_{ij}\omega^{ij}_{12},$$
(where $\omega^i_1 = g\;dy^i\wedge dx^2$, $\omega^i_2 = g\;dx^1\wedge dy^i$
and $\omega^{ij}_{12} = g\;dy^i\wedge dy^j$). The pataplectic form is

$$\Omega = d\theta = d\epsilon\wedge \omega +
\sum_{\alpha,i}dp^{\alpha}_i\wedge \omega^i_{\alpha}
+ \sum_{i<j}dp_{ij}\wedge \omega^{ij}_{12}
- \sum_{\alpha,i}{p^{\alpha}_i\over g}{\partial g\over \partial x^{\alpha}}
dy^i\wedge \omega
+ \sum_{i<j}\sum_{\alpha}p_{ij}{\partial g\over \partial x^{\alpha}}
dx^{\alpha}\wedge dy^i\wedge dy^j.$$

The Legendre correspondance is generated by the function

$$W(x,u,v,p):= \epsilon + p^{\alpha}_iv^i_{\alpha} +
p_{ij}v^i_1v^j_2 - L(x,u,v) = \epsilon + p^{\alpha}_iv^i_{\alpha} 
- {1\over 2}M^{\alpha\beta}_{ij}(x,y,p)v^i_{\alpha}v^j_{\beta},$$
where we have denoted

$$M^{\alpha\beta}_{ij}(x,y,p):=
h_{ij}(y)g^{\alpha\beta}(x)
+ \left( {b_{ij}(y)\over g(x)}-p_{ij}\right) \epsilon^{\alpha\beta}
= G^{\alpha\beta}_{ij}(x,y) -p_{ij}\epsilon^{\alpha\beta}.$$
This correspondance is given by the relation ${\partial W\over \partial v^i_{\alpha}} = 0$
which gives

\begin{equation}\label{string2}
M^{\alpha\beta}_{ij}(x,y,p)v^j_{\beta} = p_i^{\alpha}.
\end{equation}
Thus, given $(x,y,p)$, finding $(x,y,v,w)$ such that $(x,u,v,w)\leftrightarrow (x,y,p)$
amounts to solving first the linear system (\ref{string2}) for $v$ and then
$w$ is just $W(x,y,v,p)$. This system has a solution in general in the open subset
${\cal O}$ of ${\cal M}$ on which the matrix

$$M = \left( \begin{array}{cc}
h_{ij}(y)g^{11}(x) & h_{ij}(y)g^{12}(x) + {b_{ij}(y)\over g(x)}-p_{ij}\\
h_{ij}(y)g^{21}(x) - {b_{ij}(y)\over g(x)}+ p_{ij} & h_{ij}(y)g^{22}(x)
\end{array}\right)$$
is invertible.
We remark that ${\cal O}$ contains actually the submanifold
${\cal R}:= \{(x,y,p)\in {\cal M}/ g(x)p_{ij} = b_{ij}(y)\}$, so that the Legendre
correspondance induces a diffeomorphism between $T{\cal Y}\otimes T^{\star}{\cal X}$
and ${\cal R}$.\\

We shall need to define on ${\cal O}$ the inverse of $M$, i. e.
$K^{ij}_{\alpha\beta}(x,y,p)$ such that

\begin{equation}\label{string3}
K^{ij}_{\alpha\beta}(x,y,p)M^{\beta\gamma}_{jk}(x,y,p) =
\delta^i_k\delta^{\gamma}_{\alpha}.
\end{equation}
Now we can express the solution of (\ref{string2}) by

\begin{equation}\label{string4}
v^i_{\alpha} = K^{ij}_{\alpha\beta}(x,y,p)p_j^{\beta}
\end{equation}
and the Hamiltonian function is

$${\cal H}(x,y,p):= \epsilon + {1\over 2}K^{ij}_{\alpha\beta}(x,y,p)p_i^{\alpha}p_j^{\beta}.$$
We use as conjugate variables the position functions $y^i$ and the momentum
1-forms

$$P_i:= {\partial \over \partial y^i}\inn \theta
= p^{\alpha}_i\omega_{\alpha} + g\; p_{ij}dy^j.$$

The Poisson brackets are computed as follows. First, in order to obtain
$\{{\cal H}\omega, y^i\}_{\omega}$, we need to charaterize some relevant components of a
given Hamiltonian 2-vector field. Let $X$ be such a 2-vector field, writing

$$X = {\partial \over \partial x^1}\wedge {\partial \over \partial x^2}
+ X^i_1{\partial \over \partial y^i}\wedge {\partial \over \partial x^2}
+ X^i_2{\partial \over \partial x^1}\wedge {\partial \over \partial y^i}
+ \hbox{ etc }\dots ,$$
we deduce from $X\inn \Omega = d{\cal H}$ mod ${\cal I}$ that

$$g\; X^i_{\alpha} = {\partial {\cal H}\over \partial p_i^{\alpha}}.$$
Thus

$$\begin{array}{ccl}
X\sharp y^i & = & (X\inn dx^1\wedge dy^i)\omega_1 + (X\inn dx^2\wedge dy^i)\omega_2\\
 & = & \displaystyle g\;X^i_{\alpha} dx^{\alpha} =
{\partial {\cal H}\over \partial p_i^{\alpha}}dx^{\alpha}\\
 & = & K^{ij}_{\alpha\beta}p^{\beta}_jdx^{\alpha}.
\end{array}$$
Hence $y^i$ is admissible and

$$\{{\cal H}\omega, y^i\}_{\omega} = K^{ij}_{\alpha\beta}p^{\beta}_jdx^{\alpha}.$$

Next we compute $dP_i$:

$$\begin{array}{ccl}
dP_i & = & \displaystyle dp^{\alpha}_i\wedge \omega_{\alpha} + g\;dp_{ij}\wedge dy^j
+ p^{\alpha}_i{\partial g\over \partial x^{\alpha}}{\omega\over g}
+ p_{ij}{\partial g\over \partial x^{\alpha}}dx^{\alpha}\wedge dy^j\\
& = & \displaystyle - {\partial \over \partial y^i}\inn \Omega.
\end{array}$$
Hence

$$\Xi(P_i) = {\partial \over \partial y^i}.$$

We deduce that

$$\begin{array}{ccl}
\{P_i,y^j\} & = & \Xi(P_i)\inn dy^j = \delta^j_i\\
\{{\cal H}\omega,P_i\} & = & \displaystyle -\Xi(P_i)\inn d({\cal H}\omega)
= - {\partial K^{jk}_{\alpha\beta}\over \partial y^i}p^{\alpha}_jp^{\beta}_k\omega.
\end{array}$$
Notice that, because of (\ref{string3}),

$${\partial K^{jk}_{\alpha\beta}\over \partial y^i} =
- K^{jl}_{\alpha\gamma}{\partial M^{\gamma\delta}_{lm}\over \partial y^i}
K^{mk}_{\delta\beta},$$
and thus

$$\{{\cal H}\omega,P_i\} = {\partial M^{\gamma\delta}_{lm}\over \partial y^i} K^{jl}_{\alpha\gamma}K^{mk}_{\delta\beta}
p^{\alpha}_jp^{\beta}_k\omega.$$

The equations of motion are 

\begin{equation}\label{string5}
\begin{array}{ccccl}
{\bf d}y^i & = & \{{\cal H}\omega, y^i\}_{\omega} & = &
K^{ij}_{\alpha\beta}p^{\beta}_jdx^{\alpha}\\
 & & & & \\
{\bf d}P_i & = & \{{\cal H}\omega,P_i\} & = & \displaystyle
{\partial M^{\gamma\delta}_{lm}\over \partial y^i}
K^{jl}_{\alpha\gamma}K^{mk}_{\delta\beta}p^{\alpha}_jp^{\beta}_k\omega,
\end{array}
\end{equation}
along the graph $\Gamma$ of any solution of the Hamilton equations. From the first equation
we deduce that

\begin{equation}\label{string6}
{\partial y^i\over \partial x^{\alpha}} = K^{ij}_{\alpha\beta}p^{\beta}_j
\quad \Longleftrightarrow \quad
p^{\alpha}_i = M^{\alpha\beta}_{ij}{\partial y^j\over \partial x^{\beta}}.
\end{equation}
Now using (\ref{string6}) we see that along $\Gamma$,

$$\begin{array}{ccl}
P_{i|\Gamma} & = & (p^{\alpha}_i\omega_{\alpha} + g\;p_{ij}dy^j)_{|\Gamma}\\
 & = & \displaystyle \left( M^{\alpha\beta}_{ij}{\partial y^j\over \partial x^{\beta}}
+ p_{ij}\epsilon^{\alpha\beta}{\partial y^j\over \partial x^{\beta}}\right) \omega_{\alpha}\\
 & = & \displaystyle G^{\alpha\beta}_{ij}{\partial y^j\over \partial x^{\beta}}
\omega_{\alpha}
= \left( h_{ij}g^{\alpha\beta} + {b_{ij}\over g}\epsilon^{\alpha\beta}\right)
{\partial y^j\over \partial x^{\beta}} \omega_{\alpha},
\end{array}$$
and so the left hand side of the second equation of (\ref{string5}) is

$$dP_{i|\Gamma} = {1\over g}{\partial \over \partial x^{\alpha}}
\left[ g\;G^{\alpha\beta}_{ij}{\partial y^j\over \partial x^{\beta}}\right]
\omega.$$
And still using (\ref{string6}) the right hand side of the second equation of (\ref{string5})
along $\Gamma$ is

$${\partial M^{\alpha\beta}_{jk}\over \partial y^i}
{\partial y^j\over \partial x^{\alpha}}{\partial y^k\over \partial x^{\beta}}\omega.$$
Hence we recover the Euler-Lagrange equation (\ref{string1}).

\section{Conclusion}

We obtained an Hamiltonian formulation for variational problems
with an arbitrary number of variables. This could be the starting point for
building a fully relativistic quantum field theory
without requiring the space-time to be Minkowskian. This will be the subject of a
forthcoming paper, where also a $\mathfrak{p}$-bracket will be defined between
forms of arbitrary degrees.
Notice also that we may enlarge the concept of pataplectic manifolds as
manifolds equipped with a closed $n$-form and extend to this context
notions like the $\mathfrak{p}$-bracket.\\\\

{\bf{Acknowledgements}}\\
We thank G. Sardanashvily for bringing to our attention the papers \cite{Enriquez1, Enriquez2, Enriquez3, Giachetta, Sardanashvily1, Sardanashvily2, Sardanashvily3}

\end{document}